\documentclass[twocolumn,twocolappendix]{aastex631}
\pdfoutput=1 
\usepackage{amsmath,amstext}
\usepackage[T1]{fontenc}
\usepackage[figure,figure*]{hypcap}
\usepackage{mwe}
\usepackage{lipsum}
\usepackage{graphicx} 
\usepackage{gensymb}
\usepackage{multirow}
\usepackage{xcolor,colortbl}
\usepackage{tipa}
\usepackage{natbib}
\usepackage{hyperref}
\usepackage{ulem}
\usepackage{float}
\usepackage{xfrac}
\usepackage{enumitem}

\newcommand{\gaia}{\textit{Gaia}\xspace}
\newcommand{\LB}{Local Bubble\xspace}
\newcommand{\planck}{\textit{Planck}\xspace}

\newcommand{\lmax}{$l_{\rm{max}}$\xspace}

\def\pdeg{\ifmmode $\setbox0=\hbox{$^{\circ}$}\rlap{\hskip.11\wd0 .}$^{\circ} \else \setbox0=\hbox{$^{\circ}$}\rlap{\hskip.11\wd0 .}$^{\circ}$\fi}
\newcommand{\Bo}{$\textbf{B}_0$\xspace}
\newcommand{\B}{$\textbf{B}$\xspace}

\newcommand\HII{$\textrm{H}\scriptstyle\mathrm{II}$}
\newcommand{\upm}[2]{\raisebox{0.5ex}{\tiny$^{+#1}_{-#2}$}\xspace}

\newcommand{\VecPsi}{$\Vec{\psi}_\perp$\xspace}
\newcommand{\PsiPerp}{$\psi_\perp$\xspace}
\newcommand{\nside}[1]{$N_{\rm{side}} = {#1}$\xspace}
\newcommand{\Hpx}{HEALPix\xspace}
\newcommand{\dthetap}{$\Delta \theta_{\rm{planck}}$\xspace}
\newcommand{\ellbB}[2]{$(\ell_B, b_B) = ({#1}\degree, {#2}\degree)$}

\newcommand\AI{Assumption $\mathrm{I}$\xspace}
\newcommand\AII{Assumption $\mathrm{II}$\xspace}

\newcommand{\co}{$c_0$\xspace}

\newcommand{\juan}[1]{{\color{black} #1}}

\shorttitle{Local Bubble's 3D Magnetic Field}
\shortauthors{O'Neill et al.}

\begin{document}


\title{A 3D Model of the Local Bubble's Magnetic Field: Insights from Dust and Starlight Polarization}

\author[0000-0003-4852-6485]{Theo J. O'Neill}
\affiliation{Center for Astrophysics $|$ Harvard \& Smithsonian, 60 Garden St., Cambridge, MA 02138, USA}

\author[0000-0003-1312-0477]{Alyssa A. Goodman}
\affiliation{Center for Astrophysics $|$ Harvard \& Smithsonian, 60 Garden St., Cambridge, MA 02138, USA}

\author[0000-0002-0294-4465]{Juan D. Soler}
\affiliation{Istituto di Astrofisica e Planetologia Spaziali (IAPS). INAF. Via Fosso del Cavaliere 100, 00133 Roma, Italy}

\author[0000-0002-2250-730X]{Catherine Zucker}
\affiliation{Center for Astrophysics $|$ Harvard \& Smithsonian, 60 Garden St., Cambridge, MA 02138, USA}

\author[0000-0002-6800-5778]{Jiwon Jesse Han}
\affiliation{Center for Astrophysics $|$ Harvard \& Smithsonian, 60 Garden St., Cambridge, MA 02138, USA}

\begin{abstract}
Clustered stellar feedback creates expanding voids in the magnetized interstellar medium known as superbubbles.  Although theory suggests that superbubble expansion is influenced by interstellar magnetic fields, direct observational data on 3D superbubble magnetic field geometry is limited. The Sun's location inside the Local Bubble provides a unique opportunity to infer a superbubble's 3D magnetic field orientation, under the assumptions that: $\mathrm{I}$) the Local Bubble's surface is the primary contributor to plane-of-the-sky polarization observations across much of the sky, and $\mathrm{II}$) the Local Bubble's magnetic field is tangent to its dust-traced shell.  In this work, we validate these assumptions and construct a model of the Local Bubble's 3D B-field orientation from $\textit{Planck}$ 353 GHz polarization observations and a 3D-dust-derived model of the Local Bubble's shell. We test Assumption $\mathrm{I}$ by examining correlations between the Local Bubble's 3D geometry, dust polarization, and starlight polarization. We find that the Local Bubble likely dominates the polarized signal in the majority of lines of sight.  We jointly test Assumptions $\mathrm{I}$ and $\mathrm{II}$ by applying our reconstruction method to a simulated superbubble, where we successfully reconstruct the 3D magnetic field orientation over the bulk of its surface.  Finally, we use our 3D B-field model to infer the initial magnetic field orientation in the solar neighborhood prior to the Local Bubble's formation, and derive an orientation parallel to the present-day Local Arm of the galaxy. These findings provide new insights into the co-evolution of superbubbles and the magnetized interstellar medium.
\end{abstract}

\section{Introduction}

Low-density cavities in the interstellar medium (ISM) generated by feedback from sequential supernovae are ubiquitous throughout the Milky Way \citep{heiles1979,PinedaArzoumanian2023} and beyond \citep{WatkinsBarnes2023,SandstromKoch2023}.
These ``superbubbles'' are expected to play a significant role in concentrating and distributing the effects of stellar feedback \citep{KrumholzBate2014,KellerWadsley2014} and in triggering the formation of dense gas and stars \citep{Elmegreen2011,Dawson2013}.  Although theoretical models and numerical simulations predict that the orientation of magnetic fields in the ISM has a significant impact on the direction of superbubble expansion \citep[e.g.,][]{FerriereMacLow1991,Tomisaka1998, NtormousiDawson2017}, the overall role of magnetic fields in regulating the formation, expansion, and interaction of superbubbles with their environment is unclear.

While 2D measurements of superbubble magnetic field orientation and strength have been obtained \citep[e.g.,][]{WisniewskiBjorkman2007, GaoReich2015, ThomsonMcClureGriffiths2018, SolerBracco2018, JoubaudGrenier2019}, efforts to map the 3D orientations of magnetic fields over the surfaces of superbubbles have been limited by the uncertainties on their 3D geometries and by line-of-sight (LOS) confusion in the distances to structures contributing to the observed plane-of-the-sky (POS) B-field.  Efforts to probe magnetic fields in 3D through the combination of POS and LOS tracers of B-field orientation have yielded rich insights into the effects of magnetic field structure on individual molecular clouds on the surface of superbubbles \citep{TahaniLupypciw2022,TahaniGlover2022}, but has so far remained out of reach over entire superbubble surfaces.

The Local Bubble is a superbubble centered roughly near the Sun's present-day location \citep[see reviews in][]{CoxReynolds1987, WelshShelton2009, LinskyRedfield2021}.  \citet{ONeillZucker2024} recently derived a new model of the \LB's dust-traced shell from 3D dust maps, revealing the \LB as an irregular surface characterized by a prominent ``chimney'' in the high-latitude Northern hemisphere extending into the lower Galactic halo.  The exact mechanisms regulating the Local Bubble's formation and expansion into the disk and lower halo in the context of the magnetized ISM are not yet fully understood. 

Magnetic fields have likely played a key role in determining the \LB's expansion. Past work studying local starlight polarization has revealed critical insights into the likely properties of the \LB's B-field \citep[e.g.,][]{Heiles1996,SantosCorradi2011, BerdyuginPiirola2014,MedanAndersson2019,GontcharovMosenkov2019}, including that it likely makes a large contribution to the total observed sub-mm dust polarization at high latitudes \citep{SkalidisPelgrims2019}.  \citet{AlvesBoulanger2018} modeled magnetic fields on the surface of the Local Bubble at high Galactic latitudes ($|b|$\,$>$\,60$^\circ$) with the simplifying assumption that the shell has an ellipsoidal geometry, and concluded that variations in the B-field in these polar regions would be sufficient to distort measurements of Galactic B-fields by a significant amount. \citet{PelgrimsFerriere2020} extended this work with a more physically motivated geometry (enabled by 3D dust mapping) to explore the initial orientation of the local Galactic Magnetic Field in the solar neighborhood.  However, these efforts were also limited to the high-latitude regime.

A complete 3D model of the Local Bubble's magnetic field structure over all latitudes would not only be a useful probe of the local history of the magnetized ISM, but also constrain variations between the local and larger-scale Galactic magnetic fields.  In this work, we create the first complete 3D model of the Local Bubble's magnetic field.  Specifically, we constrain the Local Bubble's magnetic field structure by combining its dust-traced surface geometry with \textit{Planck} polarization observations and background starlight polarization measurements.  We assume that: $\mathrm{I})$ the observed POS magnetic field is primarily generated by a polarizing layer on the surface of the Local Bubble, and $\mathrm{II})$ the 3D magnetic field is tangent to the surface of the Local Bubble. Combining these assumptions results in a complete set of 3D B-field vector orientations.  This enables a full characterization of the B-field orientation over the \LB's surface to create the first 3D model of a superbubble's magnetic field. 

This paper is structured as follows.  We describe the data used in our analysis in \S\ref{S:obs}.  We outline the geometric construction of the 3D B-field model from 2D polarization observations in \S\ref{S:reconstruct}.  We explore various tests of \AI in \S\ref{S:assum1} through analysis of starlight and sub-mm polarization observations.  We explore tests of \AI and \AII in \S\ref{S:assum2} by applying our B-field projection method to a simulated superbubble.  We present our 3D model of the \LB's B-field in \S\ref{S:modeling}, which we use to model the initial orientation of the local Galactic magnetic field in the solar neighborhood before disruption by the \LB.  Finally, we discuss the implications of our work in \S\ref{S:discuss} before concluding in \S\ref{S:conclude}.

\section{Data}\label{S:obs}

\subsection{Dust-traced Model of the Local Bubble}

In this work, we make use of the \citet[][hereafter O24]{ONeillZucker2024} model of the \LB's dust-traced shell.  This model was constructed from the \citet{EdenhoferZucker2024} 3D map of dust extinction within 1.25 kpc of the Sun, which in turn was inferred from the \citet{ZhangGreen2023} catalog of stellar distance and extinction estimates derived from \textit{Gaia} BP/RP spectra.

The O24 model of the \LB was constructed by identifying the \LB's shell as the first ``significant'' peak in differential extinction, $A'$, along the LOS from the Sun.  The relative significance of a peak was defined using its prominence $P$ (the height of a peak above its base).  The inner and outer edges of peaks in the fiducial O24 model were defined by a half-prominence criterion ($A'_{0.5} = A'_{\rm{peak}} - 0.5 P$).  O24 also defined a more generous one-tenth prominence model ($A'_{0.9}$) for shell width; we employ both of these boundary definitions at various times in this work.  The model is sampled at \Hpx \citep{GorskiHivon2005} \nside{256} resolution.

Additional properties of the \LB's shell derived by O24 that we make use of in this work include: total extinction in \gaia G-band $A_G$, peak volume density of hydrogen $n_{\rm{peak}}$ (derived from $A'$ by assuming a fixed ratio between extinction and hydrogen column density), the inclination of the shell to the POS $\gamma$, and the normal vector to the shell in cartesian space $\textbf{n}$. 

\subsection{Planck Dust Polarization Observations}\label{S:planck_data}

To constrain the B-field in the \LB's dust-traced shell, we used the component-separated 353 GHz Stokes $I$, $Q$, and $U$ maps of Galactic dust polarization released as part of the 2018 Planck data release \citep{PlanckCollaborationAkrami2020}\footnote{\href{https://irsa.ipac.caltech.edu/data/Planck/release_3/all-sky-maps/previews/COM_CompMap_IQU-thermaldust-gnilc-unires_2048_R3.00/index.html}{COM\_CompMap\_IQU-thermaldust-gnilc-unires\_2048\_R3.00}} at uniform FWHM = 80' resolution.  The maps were produced using the Generalized Needlet Internal Linear Combination (GNILC) algorithm \citep{RemazeillesDelabrouille2011} to separate foreground Galactic polarized thermal dust emission from polarized CMB emission \citep{PlanckCollaborationXII2020}, making these data appropriate for our study of nearby Galactic polarization. 

We smoothed the maps to a full width at half maximum (FWHM) resolution of 2$\degree$.  As a result of this generous smoothing, we do not perform any debiasing.  Following \citet{PlanckCollaborationXII2020}, we correct intensity $I$ for the cosmic infrared background (-452 $\mu$K$_{\rm{CMB}}$) and Galactic zero level offset (+63 $\mu$K$_{\rm{CMB}}$).  We convert from K$_{\rm{CMB}}$ to MJy sr$^{-1}$ using the unit conversion factor 287.5 MJy sr$^{-1}$ K$^{-1}_{\rm{CMB}}$ \citep{PlanckCollaboration2020_III}.  We downsample the map from its native resolution of \Hpx \nside{2048} to \nside{256} to match the angular spacing of the \citet{ONeillZucker2024} map of the \LB. 

We calculate polarized intensity as \juan{$P$\,$=$\,$\sqrt{Q^2 + U^2}$} and polarization fraction as $p = P / I$.  We calculate polarization angle $\phi$ as
\begin{equation}
    \phi = \frac{1}{2} \arctan(-U, Q),
    \label{eqn:polangle}
\end{equation}
following IAU convention, where angles are measured relative to the North Galactic Pole with positive values increasing with increasing Galactic latitude $\ell$ \citep{HamakerBregman1996}.  The POS B-field orientation $\psi_\perp$ can be inferred from $\phi$ as
\begin{equation}
    \psi_\perp = \phi + \frac{\pi}{2}.
    \label{eqn:phi_angle}
\end{equation}
This stems from the assumption that, in the presence of an adequately strong B-field, the spinning, elongated dust grains in a given region of the ISM are likely to become preferentially aligned with their long axes perpendicular to the B-field. As a result, the thermal radiation emitting from these warm dust grains is polarized such that the observed $\phi$ is perpendicular to the POS component of the B-field $\psi_\perp$ \citep[see][for a review]{AnderssonLazarian2015}.  

Our goal in this work is to infer the 3D B-field orientation \B, which will, by construction, appear identical to the pseudovector \VecPsi when viewed in 2D projection on the POS.    

\subsection{Starlight Polarization Observations}\label{S:starlight_data}

Dust polarization observations (\S\ref{S:planck_data}) yield the projected orientation of magnetic fields averaged over an entire LOS, but convey no direct information on distance to polarizing features.  Optical polarimetry samples the LOS portion between the Sun and each star, providing additional information to resolve this ambiguity.  To this end, we supplement the \planck dust polarization information with background starlight polarimetry observations, which allow us to constrain polarization as a function of distance from the \LB's surface.

For this analysis, we identify the stars in the optical starlight polarization catalog assembled by \citet{PanopoulouMarkopoulioti2025} for which the \LB contributes \juan{most} of the integrated extinction along the LOS.  To perform this cut, we restrict our analysis to stars with median \textit{Gaia} EDR3 photogeometric distances \citep{BailerJonesRybizki2021} within the boundaries of the \citet{EdenhoferZucker2024} 3D dust map, \juan{$d_\star$\,$<$\,$1250$\,pc}. We calculated the ratio of the integrated extinction from the Sun to the star, as defined in the \citet{EdenhoferZucker2024} dust map, to the extinction from the Sun to the outer edge of the \LB,
\begin{equation}
    R_\star = \frac{A_\star}{A_{LB,outer}} = \frac{\sum_0^{d_\star} A_{ZGR23}}{\sum_0^{d_{LB, outer}} A_{ZGR23}}
\end{equation}
We selected stars with $R_\star < 2$, i.e., stars whose foreground extinction are at most twice the extinction associated with the \LB's interior and shell.  We additionally excluded stars identified by the \citet{PanopoulouMarkopoulioti2025} catalog as possessing intrinsic polarization.  Finally, we required all observations to have measurements of polarization fraction $p$, polarization fraction uncertainty $e_p$, and polarization angle $\phi$, and exclude measurements with extremely low $p$ ($p < 0.01 \%$).  

In total, this yielded a sample of 10,795 stars with polarization fraction and orientation measurements.\footnote{The individual publications whose data were included in our analysis are, in order of number of measurements used:  \citet{Heiles2000AJ....119..923H,Franco2015ApJ...807....5F,BerdyuginPiirola2014,Santos2014ApJ...783....1S,SantosCorradi2011,Pereyra2004ApJ...603..584P,Lobo2015ApJ...806...94L,Panopoulou2015MNRAS.452..715P,Targon2011ApJ...743...54T,Moneti1984ApJ...282..508M,Vrba1976AJ.....81..958V,Goodman1990ApJ...359..363G,Berdyugin2002AA...384.1050B,Alves2007AA...470..597A,Weitenbeck2008AcA....58..433W,Berdyugin2001AA...372..276B,PanopoulouTassis2019,Zejmo2017MNRAS.464.1294Z,Heyer1987ApJ...321..855H,Vaillancourt2020ApJ...905..157V,Alves2006,Andersson2010ApJ...720.1045A,Soam2015AA...573A..34S,Singh2022MNRAS.513.4899S,Slowikowska2018MNRAS.479.5312S,Oudmaijer2001AA...379..564O,BerdyuginTeerikorpi2001AA...368..635B,Neha2018MNRAS.476.4442N,Soam2017MNRAS.465..559S,Alves2014AA...569L...1A,Eswaraiah2011MNRAS.411.1418E,Wang2017ApJ...849..157W,Poidevin2006ApJ...650..945P,Topasna2020PASP..132d4301T,Santos2012ApJ...751..138S,Andersson2007ApJ...665..369A,Whittet2001ApJ...547..872W,Piirola2020AA...635A..46P,Eswaraiah2019ApJ...875...64E,Medhi2010MNRAS.403.1577M,Cotton2019MNRAS.483.3636C,Weitenbeck2008AcA....58...41W,Bailey2010MNRAS.405.2570B,Panopoulou2019AA...624L...8P,Alves2011AJ....142...33A,Gil-Hutton2003MNRAS.345...97G,Soam2013MNRAS.432.1502S,Sen2000AAS..141..175S,Eswaraiah2013AA...556A..65E,Das2016ApSS.361..381D,Cotton2017MNRAS.467..873C,Pereyra2002ApJS..141..469P,Choudhury2022RAA....22g5003C,Bijas2022MNRAS.515.3352B,Pandey2013ApJ...764..172P,Andersson2013ApJ...775...84A,Choudhury2019MNRAS.487..475C,Neha2016AA...588A..45N,Berdyugin2004AA...424..873B,Chakraborty2014MNRAS.442..479C,Seron2016MNRAS.462.2266S}.}  Distances from the Sun to the background stars in our sample range from $d = 7 \-- 1250$ pc, encompassing stars both interior and exterior to the \LB's surface. 

Polarization fraction $p$ is known to be a biased estimator of linear polarization strength.  Following \citet{PanopoulouMarkopoulioti2025}, we estimated debiased polarization fractions $p_d$ as \citep{PlaszczynskiMontier2014},
\begin{equation}
    p_{\rm{d}} = p - e_p^2 \frac{1 - e^{-p^2 / e_p^2}}{2p}.
\end{equation}

Polarization angles $\phi$ in the \citet{PanopoulouMarkopoulioti2025} catalog are derived in an equatorial frame, increasing to the east from the North Celestial Pole and ranging between [-90$\degree$, 90$\degree$].  We converted the polarization angles to a Galactic frame, increasing to the east from the North Galactic Pole, following \citet{PanopoulouTassis2016} (assuming a J2000 frame where the North Celestial Pole is located at ($\ell_{\mathrm{n}}$, $b_{\mathrm{n}}$) = (122.93$\degree$, 27.13$\degree$)).  For a given measurement with declination $\delta$ and Galactic latitude $\ell$, Galactic polarization angle $\phi_G$ can be derived from its equatorial polarization angle $\phi_E$ as  
\begin{equation}
 \phi_\star = \phi_G = \phi_E + \arctan \left[ \frac{\sin(\ell_n - \ell)}{\tan(b_n) \cos(b) - \sin(b)\cos(\ell_n-\ell)} \right].
\end{equation}
We transform $\phi_\star$ so that $\phi_\star \in [0\degree, 180\degree]$.

As discussed in the previous subsection, the long axis of dust grains tends to be preferentially aligned perpendicular to the surrounding magnetic field.  When initially unpolarized starlight encounters these aligned dust grains, a slightly larger fraction of the light will be blocked along the long axis of the grain than the short axis.  This causes the starlight itself to become polarized, such that the overall measured polarization along the LOS towards the background star is the density-weighted average of polarization imparted by the dusty ISM along the LOS.  Starlight polarization angles $\phi_\star$ are then parallel to the POS component of the B-field $\psi_\star$, 
\begin{equation}
    \psi_\star = \phi_\star,
\end{equation}
and can be compared directly to POS B-field orientations derived from dust polarization, $\psi_\perp$.

\section{Method for Reconstructing 3D B-field}\label{S:reconstruct}

\subsection{Guiding Assumptions}\label{S:assumptions}

We derived the \LB's 3D magnetic field orientation \B under the assumptions that:

\begin{enumerate}[label=\Roman*.]
    \item The \LB's surface is the last significant contributor to polarization measurements over most of the sky.
    \item \B is tangent to the dust-traced surface of the \LB.
\end{enumerate}

As we will discuss, we expect that both of these assumptions vary in accuracy and relevance over the surface of the Bubble.

Assumption $\textrm{I}$ reflects our expectation that the low-density interior of the \LB is unlikely to contain much polarizing  material, meaning that, for emission originating at infinite distance and traveling towards the solar system, the Bubble's shell is likely the last significant source of polarization along the LOS.  The Bubble's interior has been observed to contribute relatively little polarization to background sources \citep[e.g.,][]{GontcharovMosenkov2019}, and the expanding shell is likely to make an oversized contribution to the total observed dust polarization, especially at intermediate to high latitudes \citep[e.g.,][]{SkalidisPelgrims2019}.  This behavior is borne out in simulations; e.g., \citet{MaconiSoler2023} created synthetic 353 GHz dust polarization maps from an observer inside a Local Bubble-type cavity, and found that their simulated bubble acts as a polarizing ``filter'' over much of the simulated observer's sky.

Assumption $\textrm{II}$ stems from the theorized and observed behaviors of superbubbles interacting with magnetic fields. Simulations predict that in the process of a feedback-driven bubble's expansion (whether that feedback is supernova-driven on the scale of superbubbles, or stellar wind-driven on the scale of \HII{} regions), gas, dust, and magnetic fields in the surrounding ISM will be swept up into the bubble's shell, yielding magnetic fields that are tangent to the bubble's surface \citep{FerriereMacLow1991, Tomisaka1998, deAvillezBreitschwerdt2005, StilWityk2009, vanMarleMeliani2015, NtormousiDawson2017}. This behavior is held out in 2D observations of bubbles; bubble-like morphologies are frequently traced by tangent magnetic fields \citep[e.g.,][]{SolerBracco2018,BraccoBresnahan2020,TahaniBastien2023}. In their previous studies of the \LB's magnetic field, \citet{AlvesBoulanger2018} and \citet{PelgrimsFerriere2020} made this assumption of tangency while modeling the B-field at polar latitudes ($|b| \geq 60 \degree$).

We expect that \AI will be least reliable at low latitudes, i.e., in the Galactic plane where more confusion exists along the line of sight, and for LOS overlapping with low-density regions of the the \LB's shell. We expect that the geometrical constraint imposed by \AII will be least reliable in low-density regions of the shell where uncertainties on shell inclination are high \citep{ONeillZucker2024}.

\subsection{Geometric Construction}\label{S:geometric}

\begin{figure*}
    \centering
    \includegraphics[width=0.8\textwidth]{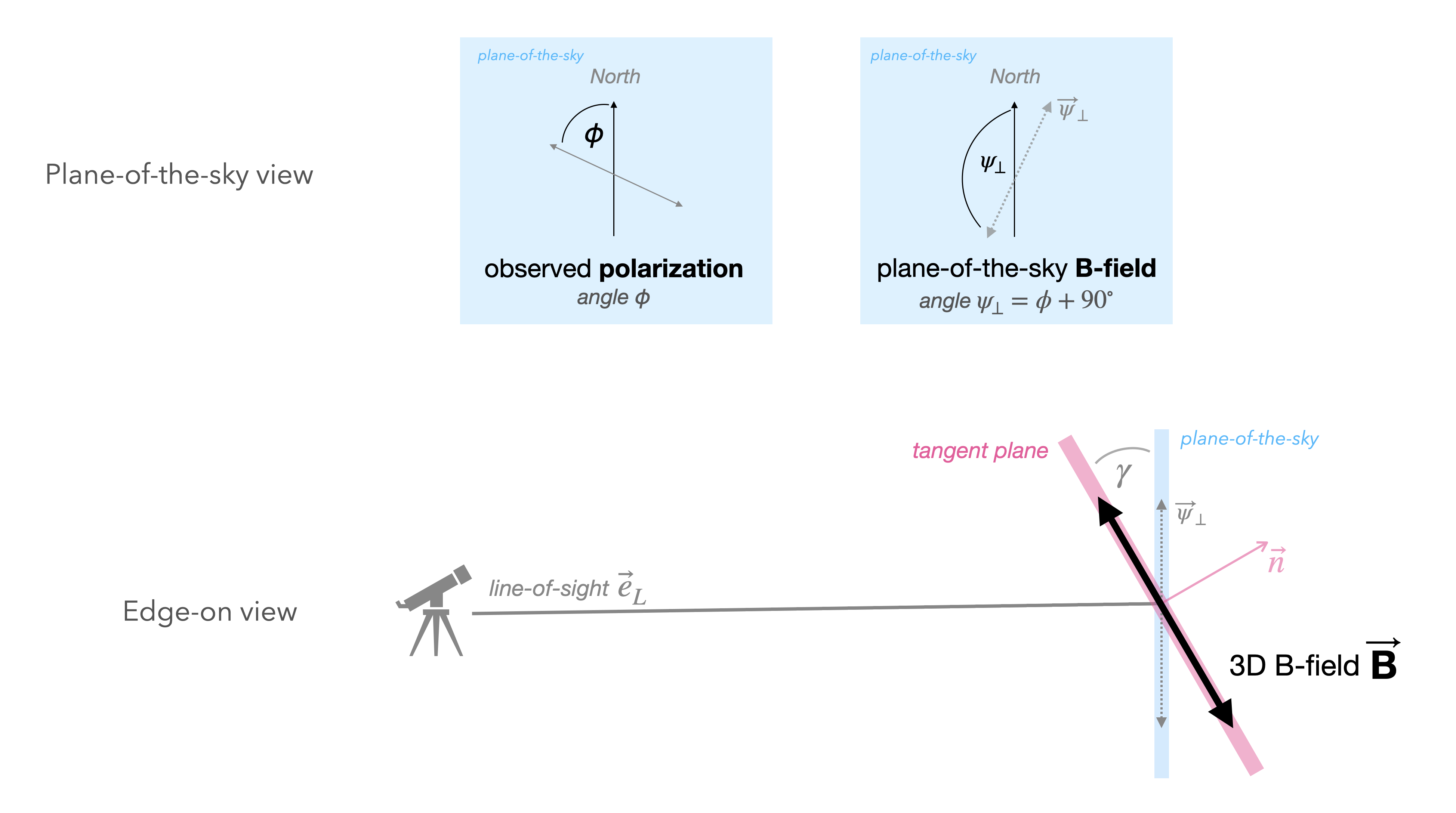}
    \caption{A schematic diagram showing the relationship between 2D and 3D magnetic fields adopted in this work.  \textit{Top row:} Plane-of-the-sky views (marked by blue squares) of an observed dust polarization vector (left, solid gray vector) and inferred B-field vector (right, dotted gray vector).  \textit{Bottom row:} An edge-on view of the adopted 3D geometry of the \LB's B-field.  The 3D magnetic field (solid black vector) is constrained to the tangent plane to the \LB's surface (thick pink line).  That vector projects to the 2D vector \VecPsi (dashed gray vector) observed in the plane-of-the-sky (thick blue line).  The tangent plane is inclined with respect to the plane-of-the-sky by \juan{angle $\gamma$}, and is characterized by its normal vector $\textbf{n}$ (pink vector).}
    \label{fig:vector_diagram}
\end{figure*}

\begin{figure*}
    \centering
    \includegraphics[width=\textwidth]{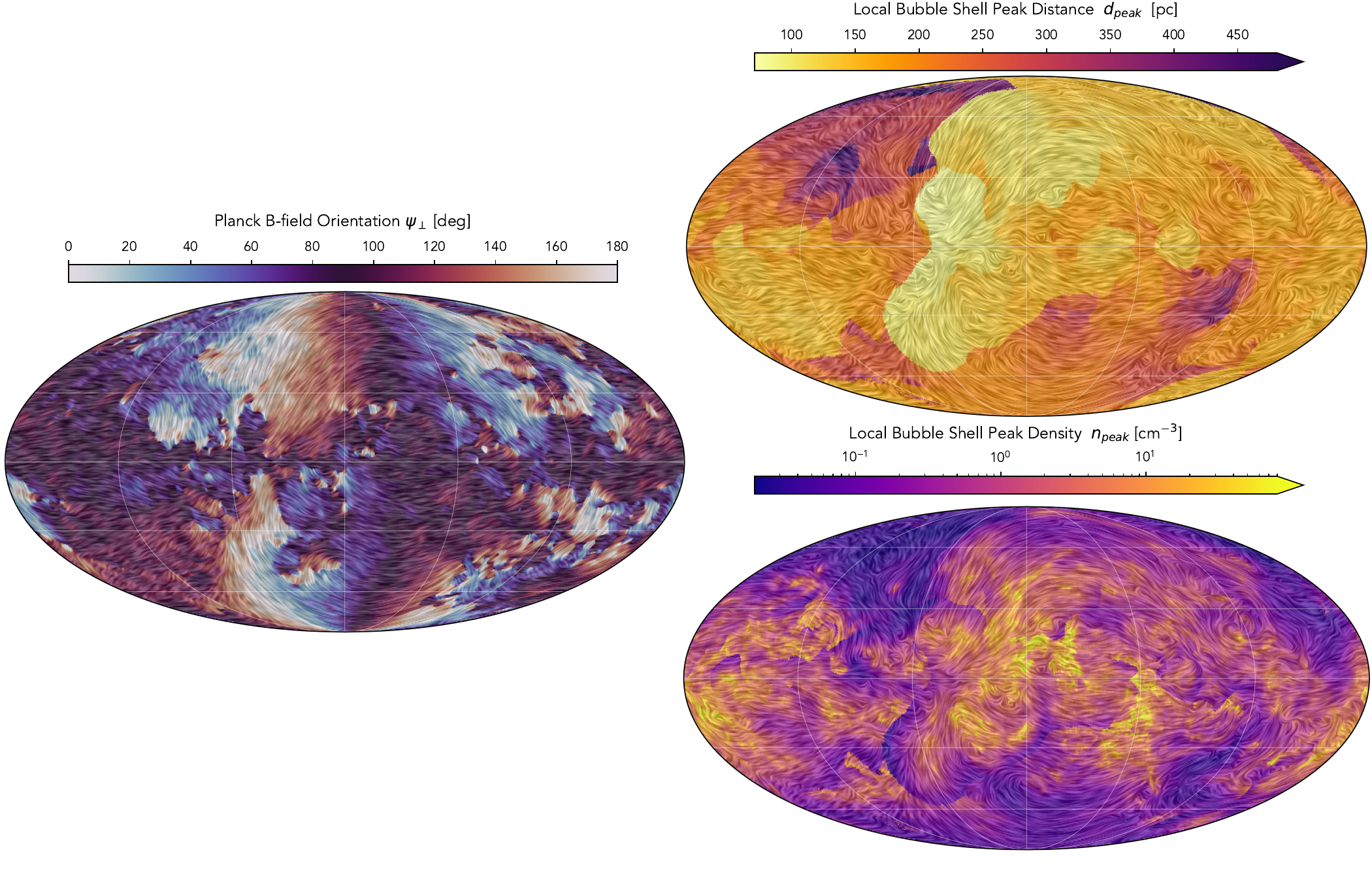}
    \caption{\textit{Left}: Mollweide projection of \planck B-field orientation, \PsiPerp, is shown, with corresponding Line Integral Convolution of the B-field orientation shown by the overlaid translucent drapery pattern.  \textit{Top right:} as left, but showing distance to the \LB's peak-density surface.  \textit{Bottom right:} as left, but showing peak density of the \LB's shell. All Mollweide projections shown throughout this work are centered towards the Galactic center at $\ell = 0^\circ$.  }
    \label{fig:planck_orientation_corr}
\end{figure*}

\begin{figure}
    \centering
    \includegraphics[width=0.47\textwidth]{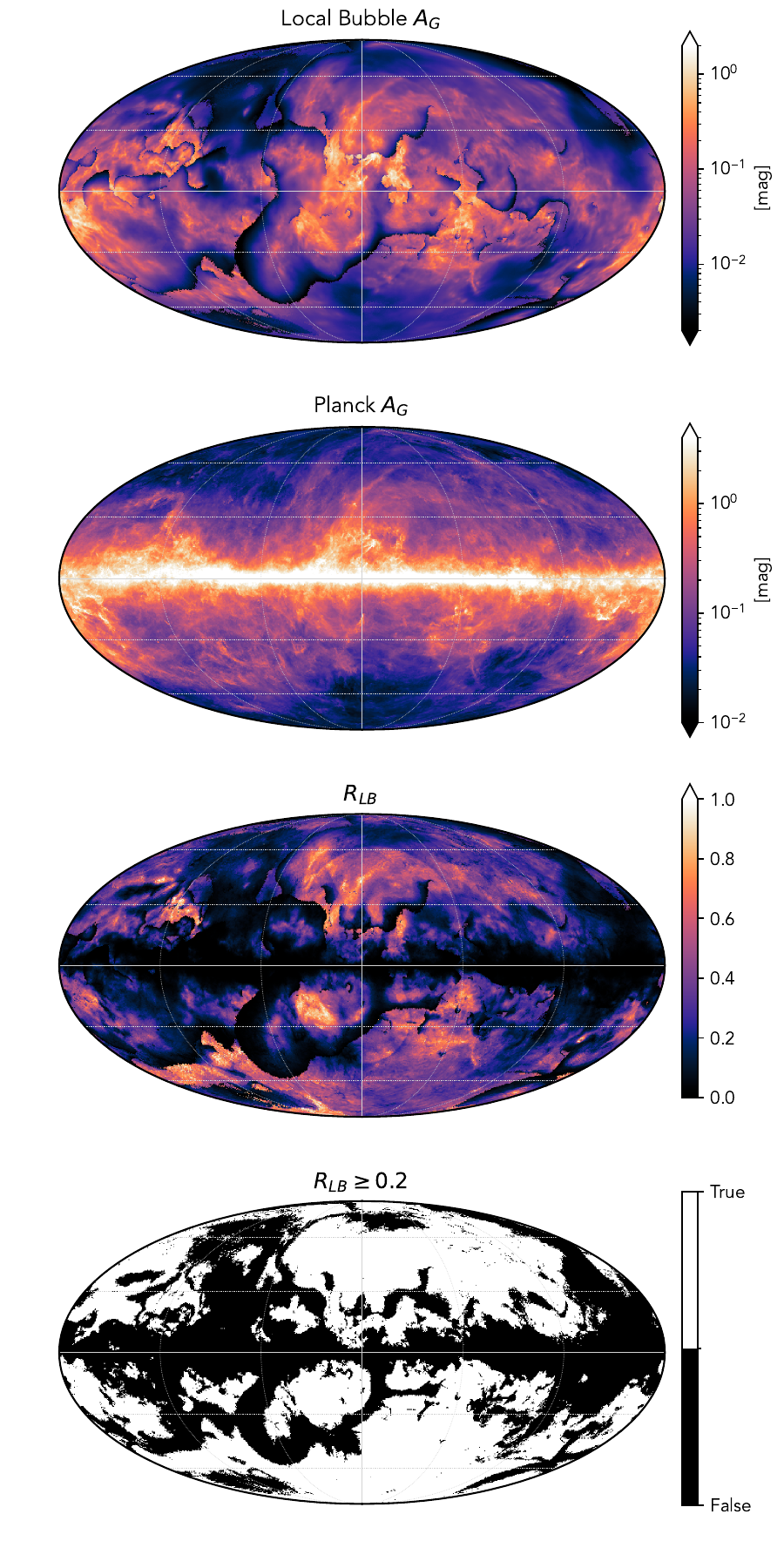}
    \caption{\textit{Top:} Integrated dust extinction contributed by the \citet{ONeillZucker2024} model of the \LB's shell, $A_{G, LB}$.  \textit{Second from Top:} Estimated total extinction derived from the \citet{PlanckInt2016_XLVIII} GNILC map of Galactic foreground reddening, $A_{G, Planck}$.  \textit{Third from Top:} Ratio, $R_{\rm{LB}}$, of $A_{G, \rm{LB}}$ to $A_{G, Planck}$.  \textit{Bottom:} Binary mask of regions with $R_{\rm{LB}} \geq 0.2$, with regions passing this criterion shown in white.}
    \label{fig:ratio_extinc}
\end{figure}

We adopt the geometry shown in \juan{Fig.~}\ref{fig:vector_diagram} to unify the true, 3D orientation of the magnetic field \B with its observed, 2D orientation projected into the POS.  In this scenario, an observer within a magnetized bubble observes a 2D POS polarization orientation $\phi$ that translates to a (perpendicular) 2D B-field pseudovector \VecPsi.  This pseudovector is a 2D projection (in Galactic coordinates, in the POS) of the true 3D vector \B.  \AII ensures that there is only one possible 3D vector orientation that could create that projected 2D pseudovector that in 3D space lies within the Bubble's tangent plane.  

We can describe this situation mathematically as follows.  Our 3D B-field is defined in a Cartesian coordinate system,
\begin{equation}
    \textbf{B}(x,y,z) = (B_x, B_y, B_z),
\end{equation}
with x pointing from the Sun towards the Galactic center at $\ell$ = 0$\degree$ [$\textbf{B} = (1, 0, 0)$], y towards $\ell$\juan{\,$=$\,}90$\degree$  [$\textbf{B} = (0, 1, 0)$], and z towards the North Galactic Pole at $b$ = 90$\degree$  [$\textbf{B} = (0, 0, 1)$].\footnote{Unless otherwise specified, all vectors $\textbf{v}$ reported in this work are unit vectors ($|\textbf{v}|=1$).}

For each point on the \LB's surface, we define the LOS as a unit vector from the Sun to the point,
\begin{equation}
    \textbf{e}_L = \left(\frac{x}{d}, \frac{y}{d}, \frac{z}{d} \right),
\end{equation}
where $d = (x^2 + y^2 + z^2)^{1/2}$.   
The tangent plane to each point on the \LB's surface was fit by O24 and is defined by its normal unit vector,
\begin{equation}
    \textbf{n} = (n_{x}, n_{y}, n_{z}).
\end{equation}
Since $\textbf{e}_L$ is the normal vector to the POS, the angle between the POS and the tangent plane to the \LB's surface can be defined as,
\begin{equation}
    \gamma = \arccos (\textbf{n} \cdot \textbf{e}_L), \rm{\ \ } \gamma \in [0\degree,90\degree].
\end{equation}
Under \AII, this is also the angle between the POS and the 3D B-field.

We matched LOS in the O24 \LB model and the \planck measurements to assign a projected B-field angle \PsiPerp to each point on the Bubble's surface.  We then created a small offset from ($\ell$, $b$) in the direction of \VecPsi to define a new position ($\Delta \ell, \Delta b$) in the POS along the projected orientation of the 2D B-field,\footnote{For a sufficiently small separation $s$, the offsets are \textit{almost} equivalent to:
\begin{equation}
\begin{split}
B_{\perp \ell} & = \sin({\psi_{\perp}})/s = \cos{(\phi)}/s \\ 
B_{\perp b} & = \cos{(\psi_{\perp})}/s = - \sin{(\phi)}/s.
\end{split}
\end{equation}
However, to account for the effects of spherical geometry, we calculated ($B_{\perp \ell}, B_{\perp b}$) using the astropy function {\tt directional\_offset\_by} \citep{astropy_2018}.}
\begin{equation}
\begin{split}
    \Delta \ell & = \ell + B_{\perp \ell}(\psi_\perp) \\
    \Delta b & = b + B_{\perp b}(\psi_\perp). 
\end{split}
\end{equation}
We transformed the offset 2D coordinates ($\Delta \ell, \Delta b$) back to 3D Cartesian coordinates, and defined a unit vector from the Sun to the 3D offset,
\begin{equation}
    \Delta \textbf{B}_\perp(\psi_\perp) 
      = (\cos \Delta \ell \cos \Delta b, \sin\Delta \ell \cos\Delta b, \sin \Delta b).
\end{equation}

We can then define the 3D B-field $\textbf{B}$ as the normal to the plane containing both 1) the normal to the \LB surface $\textbf{n}$ and 2) the normal to the plane containing $\textbf{e}_L$ and $\Delta \textbf{B}_\perp (\psi_\perp)$, 
\begin{equation}
     \textbf{B} = \textbf{B}(\textbf{n}, \psi_\perp) = \frac{\textbf{n} \times (\Delta \textbf{B}_\perp \times \textbf{e}_L)}{| \textbf{n} \times (\Delta \textbf{B}_\perp \times \textbf{e}_L) |},
     \label{eqn:bfield3d}
\end{equation}
i.e., as \juan{a unit} vector at the intersection of the \LB's tangent plane and the projection of \PsiPerp into the POS.  In other words, under Assumptions $\mathrm{I}$ and $\mathrm{II}$, $\textbf{B}$ is the 3D magnetic field orientation on the surface of the \LB.  


\section{Assumption $\textrm{I}$: The Local Bubble as the Last Significant Contributor to Polarization Metrics}\label{S:assum1}

If \AI holds, we might expect the properties of the \LB's shell to affect the amount and orientation of polarization in the \planck 353 GHz maps and in background starlight polarization observations.  In this section, we explore various correlations between the \LB and \planck (\S\ref{S:a1_planck}) and starlight (\S\ref{S:a1_starlight}) polarization.       

\subsection{Signatures of the Local Bubble in Planck Polarization Metrics}\label{S:a1_planck}

In this section, we analyze correlations between the \LB's 3D surface and 2D \planck polarization metrics.  We first consider the qualitative correlation between properties of the \LB's shell and \planck B-field orientation.  Figure \ref{fig:planck_orientation_corr} compares, in 2D projection, \planck B-field orientation with the distance to the \LB's shell and the peak density within the shell.  We observe a variety of apparent morphological correlations between dust features on the \LB's surface and the orientation of the \planck B-field, especially between higher-density features tracing local molecular clouds and filaments on the \LB's surface.  This correlation is not unexpected; morphological agreement between dust features and the \planck B-field orientation are well-established, both on the scales of individual molecular clouds \citep{PlanckCollaborationAde2016_XXXV} and on the scales of large filaments and ridges visible in all-sky maps \citep{PlanckCollaborationAdam2016_XXXII}.

We next consider quantitative correlations between the \LB's shell and \planck metrics including total intensity $I$, polarized intensity $P = \sqrt{Q^2 + U^2}$, and polarization fraction $p = P/I$.  We adopt a simple model for total signal in the \planck maps as the sum of \juan{the} signal from the \LB and the ``background''\juan{, defined as} all emission \juan{from the volume beyond} the \LB's surface),
\begin{equation}
\begin{split}
    I_{\rm{tot}} & = I_{\rm{LB}}+ I_{\rm{bkg}} \\
    Q_{\rm{tot}} & = Q_{\rm{LB}} + Q_{\rm{bkg}} \\ 
    U_{\rm{tot}} & = U_{\rm{LB}} + U_{\rm{bkg}}. \\
\end{split}
\label{eqn:IQU_tot}
\end{equation}

Following the derivation of \citet{PlanckCollaboration2015_XX}  \citep[which builds on the work of][and others]{LeeDraine1985,WardleKonigl1990} for polarization parameters integrated along the LOS (parameterized here by $dL$), we can consider emission from the \LB's surface as being described by, 
\begin{equation}
\begin{split}
    I_{\rm{LB}} & = \int  n_{\rm{LB}} \beta \left[1 - p_0 \left(\cos^2(\gamma_{\rm{LB}}) - 2/3 \right)\right]   dL \\
    Q_{\rm{LB}} & = \int n_{\rm{LB}} \beta p_0  \cos^2(\gamma_{\rm{LB}}) \cos (2 \phi_{\rm{LB, HPX}}) dL\\
    U_{\rm{LB}} & = \int n_{\rm{LB}} \beta p_0   \cos^2(\gamma_{\rm{LB}}) \sin (2 \phi_{\rm{LB, HPX}})  dL \\
    P_{\rm{LB}} &  = \int n_{\rm{LB}} \beta p_0  \cos^2(\gamma_{\rm{LB}}) dL \\
    p_{\rm{LB}} & = \frac{p_0 \cos^2(\gamma_{\rm{LB}})}{\left[1 - p_0 (\cos^2(\gamma_{\rm{LB}}) - 2/3)\right]},
\end{split}
\end{equation}
where $n_{\rm{LB}}$ is the volume density of hydrogen in the \LB's shell, $\gamma_{\rm{LB}}$ is the inclination of the \LB's B-field to the POS, $\phi_{\rm{LB, HPX}}$ is the \LB's B-field's polarization angle in \Hpx convention (as opposed to the IAU convention used in the rest of this work), and $p_0$ is a factor related to the intrinsic polarization fraction and average dust grain geometry.  For convenience, we have defined $\beta$\,$=$\,$\sigma_H B_\nu(T)$, where $\sigma_H$ is the average dust cross section per hydrogen atom at 353 GHz and $B_\nu(T)$ is the Planck function at 353 GHz and temperature $T$. 

We are interested in testing correlations that exist between these properties while holding $\beta$ and $p_0$ constant.  If we assume that $\gamma_{\rm{LB}}$ and $\phi_{\rm{LB, HPX}}$ are constant along each LOS within the \LB's shell, the integrals listed above reduce to a list of constants multiplied by $\int n_{\rm{LB}} dL = N_H$, which would be easily obtained from the O24 model's integrated extinction within each LOS of the shell by assuming a fixed conversion between extinction and column density.  However, since \juan{the} integrated density of the shell strongly depends on the convention used to define its boundaries (e.g., $A'_{0.5}$ vs $A'_{0.9}$), we prefer to consider only the peak density of dust within each LOS, $n_{\rm{peak}}$, which is invariant under shell definition convention.

In terms of testable predictions, we then investigate the following correlations:
\begin{equation}
\begin{split}
    I & \propto n_{\rm{peak}} \\
    P & \propto n_{\rm{peak}} \\
    I / n_{\rm{peak}} & \propto \left[1 - p_0 (\cos^2(\gamma_{\rm{LB}}) - 2/3)\right] \\
    P / n_{\rm{peak}} & \propto \cos^2(\gamma_{\rm{LB}})  \\
    p & \propto \frac{\cos^2(\gamma_{\rm{LB}})}{\left[1 - p_0 (\cos^2(\gamma_{\rm{LB}}) - 2/3)\right]}.
\end{split}
\label{eqn:pred_planck_corr}
\end{equation}
We adopt a uniform value of \juan{$p_0$\,$=$\,$0.2$} \citep{PlanckCollaboration2015_XX} but verify that the derived correlation coefficients are insensitive to this choice.

Under the formulation of Eqn. \ref{eqn:IQU_tot}, for quantities dependent on density (e.g., $I$) we expect the total signal to be dominated by the \LB in LOS where the integrated extinction of the \LB's shell is a significant fraction of the integrated extinction of the background signal.  In other words, we can consider the ratio of extinction from the \LB's shell to total extinction along the LOS,
\begin{equation}
    R_{\rm{LB}} = \frac{A_{\rm{LB}}}{A_{\rm{tot}}} = \frac{A_{\rm{LB}}}{A_{\rm{LB}}+A_{\rm{bkg}}},
    \label{eqn:RLB}
\end{equation}
as a proxy for LOS where we expect to have the best chance of detecting the correlations outlined in Eqn. \ref{eqn:pred_planck_corr} in the \planck maps. 

The extinction of the \LB shell in \gaia G-band $A_{G,\rm{LB}}$ was derived by \citet{ONeillZucker2024} for multiple definitions of the inner and outer boundaries of the \LB's shell;  to encompass the bulk of the shell's dust, we use $A_{G,\rm{LB}}$ derived for their supplementary definition of the shell at a 1/10th prominence criterion ($A'_{0.9}$). For total extinction along the LOS, we query the \citet{PlanckInt2016_XLVIII} GNILC map of $E(B-V)$ using the python package {\tt dustmaps} \citep{M_Green_2018}.  We convert these reddening estimates to G-band extinction assuming a fixed reddening vector of $R_V=3.1$ \citep{CardelliClayton1989} and a ratio between G and V band extinctions of $A_V/A_G = 1.36$ \citep{ZhangGreen2023}, leading to a final relationship $A_{G, \rm{tot}} = 2.28 E(B-V)$.

Figure \ref{fig:ratio_extinc} shows, in 2D Mollweide projections, the \LB shell extinction $A_{\rm{LB}}$, total \planck extinction $A_{\rm{tot}} = A_{planck}$, and their ratio $R_{\rm{LB}}$.  We define a mask based on a minimum threshold of $R_{\rm{LB}} \geq 0.2$ (the median of the all-sky distribution, equivalent to $A_{\rm{LB}} \geq 0.25 A_{\rm{bkg}}$; this mask is also shown in Figure \ref{fig:ratio_extinc}) within which to investigate correlations between \planck metrics and the \LB.  This mask largely excludes the Galactic plane, portions of the low-density Local Chimney, and the diffuse edges of \juan{higher-density} clouds within the \LB's shell. 

\subsubsection{Correlations with Shell Density}

\begin{figure*}
    \centering
    \includegraphics[width=\textwidth]{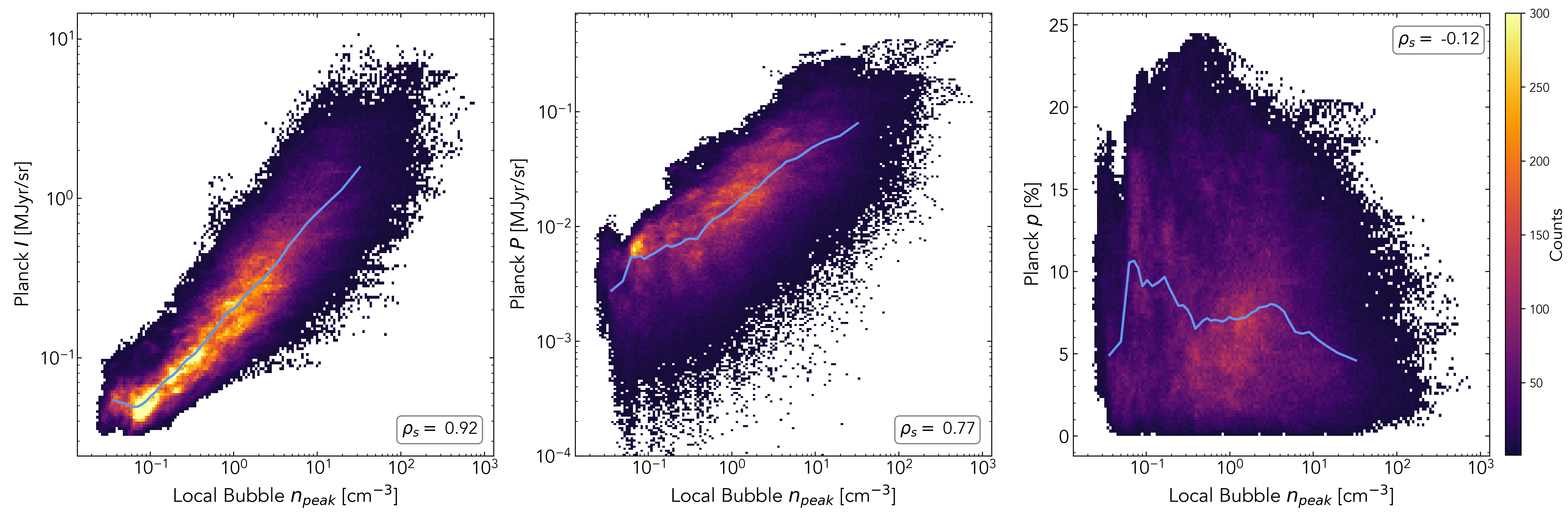}
    \includegraphics[width=\textwidth]{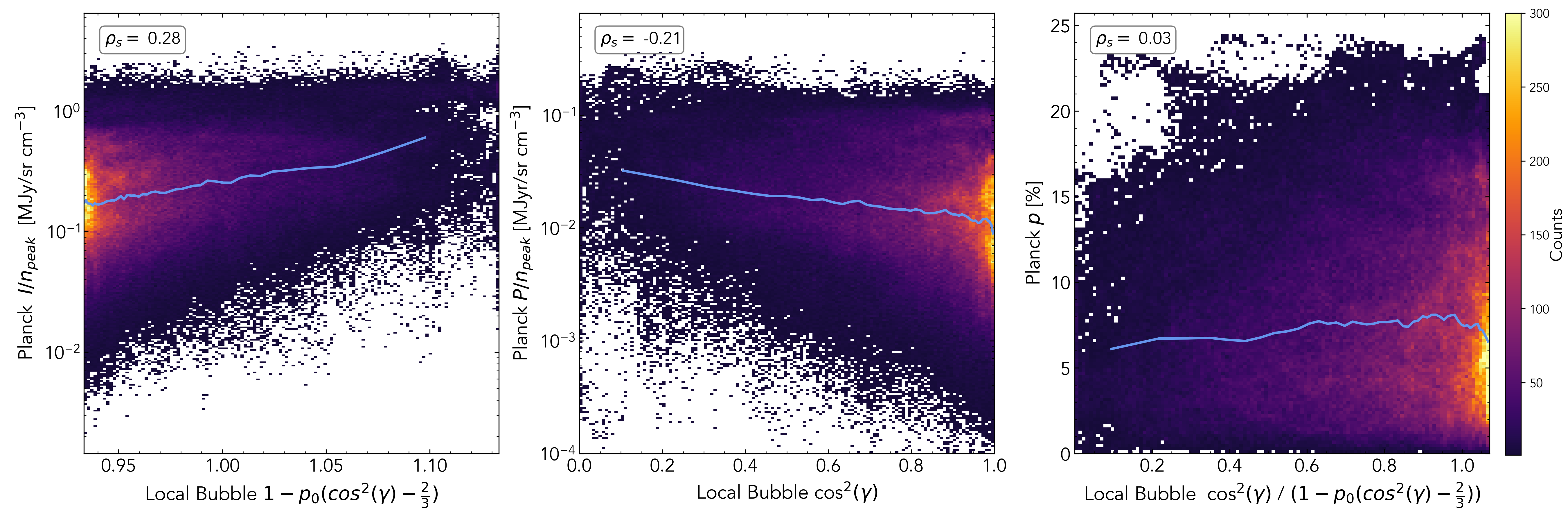}
    \caption{2D histograms showing the relationship between \planck polarization metrics and properties of the \LB's shell, in the subset of LOS with $R_{\rm{LB}} \geq 0.2$.  \textit{Top row:} Peak \LB shell density $n_{\rm{peak}}$ vs. (\textit{from left}) \planck $I$, $P$, and $p$.  \textit{Bottom row:} Terms dependent on \LB shell inclination to the POS, $\gamma$, vs. (\textit{from left}) $I/n_{\rm{peak}}$, $P/n_{\rm{peak}}$, and $p$.  In each figure, the blue line shows the median value of the independent variable as a function of \juan{the} dependent variable, in 1\% percentile bins of the dependent variable between its 1st and 99th percentile values.  Spearman correlation coefficients $\rho_s$ are reported in inset text boxes.  }
    \label{fig:planck_corr}
\end{figure*}

Our first quantitative inquiry is into correlations (or lack thereof) between peak \LB shell density $n_{\rm{peak}}$ and \planck $I$, $P$, and $p$.  As outlined in Eqn. \ref{eqn:pred_planck_corr}, we expect positive correlations between $n_{\rm{peak}}$, $I$, and $P$, and no correlation between $n_{\rm{peak}}$ and $p$.  Two-dimensional histograms of these relationships (within our mask of $R_{\rm{LB}}\geq 0.2$) are shown in Figure \ref{fig:planck_corr}.  

As expected, we observe a strong positive correlation between $n_{\rm{peak}}$ and $I$ (Spearman $\rho_s = 0.92$, $p < 0.001$) and a strong positive correlation between $n_{\rm{peak}}$ and $P$ ($\rho_s = 0.77$, $p < 0.001$).  These results are not terribly surprising: as the LOS passes through more dense material, intensity and polarized intensity both increase proportionally to the LOS density.
There is a negligible negative correlation between $n_{\rm{peak}}$ and $p$ ($\rho_s = -0.12$, $p < 0.001$).  The lack of a meaningful correlation here is expected under Eqn. \ref{eqn:pred_planck_corr}.  We note that the overall shape of this correlation is nearly identical to the previously reported correlations between $p$ and total hydrogen column density $N_H$ by \citet{PlanckCollaborationXIX2015} and  \citet{PlanckCollaborationXII2020}; this is also unsurprising, as \LB shell density directly contributes to total hydrogen column density along the LOS.  The decrease in the maximum $p$ for increasing $N_{H}$ has previously been interpreted by \citet{PlanckCollaboration2015_XX} and \citet{PlanckCollaborationXIX2015} as the result of turbulence and depolarization driven by the presence of variably-polarized structures along the LOS.

\subsubsection{Correlations with Shell Inclination}

We next consider correlations between \planck metrics and the inclination of the \LB's shell to the POS, $\gamma$.  
Figure \ref{fig:planck_corr} shows 2D histograms of the correlations between various terms dependent on $\gamma$ with $I/n_{\rm{peak}}$, $P/n_{\rm{peak}}$, and $p$ (as outlined in Eqn. \ref{eqn:pred_planck_corr}). 

As expected, a weak positive correlation exists between $I/n_{\rm{peak}}$ and $1 - p_0 (\cos^2(\gamma) - \frac{2}{3})$ ($\rho_s = 0.28$, $p < 0.001$).  In the reverse of our expectations, a weak negative correlation exists between $P/n_{\rm{peak}}$ and $\cos^2(\gamma)$ ($\rho_s = -0.21$, $p < 0.001$).  Finally, no meaningful correlation exists between $p$ and $\cos^2(\gamma) / (1 - p_0 (\cos^2(\gamma) - \frac{2}{3})$ ($\rho_s=0.03$, $p < 0.001$) (although we note that the maximum $p$ for a given $\gamma$-derived term increases with $\gamma$-derived term, as expected).  

In Appendix \ref{S:sim_corr_planck}, we evaluate the same correlations outlined in Eqn. \ref{eqn:pred_planck_corr} for a simulated \LB equivalent.  We observe remarkably similar qualitative relationships and quantitative correlation strengths as to what we have found for the real \LB.  This includes the negative correlation between $P/n_{\rm{peak}}$ and the $\gamma$ term, and the lack of correlation between $p$ and the $\gamma$ term.

The strength of any correlation between $\gamma$ and polarization metrics such as $p$ will be significantly affected by factors such as turbulence and confusion along the LOS, which we cannot directly control for in our observational data.  \citet{HalalClark2024} recently suggested that the inclination of the \LB's shell to the LOS/POS has no significant effect on \planck 353 GHz polarization fraction ($p$), and that $p$ is instead significantly influenced by the complexity of structures along a given LOS contributing to the total 3D dust distribution.  They argued that the lack of a meaningful correlation between $p$ and shell inclination (in both the \citet{PelgrimsFerriere2020} and O24 models of the \LB) indicates that either \AI and/or \AII are not physically motivated.  The lack of correlation between $p$ and $\gamma$ for the simulated Bubble that we analyze in this work (where our analysis indicates that \AI and \AII are met for the majority of LOS) suggests that the strength of the correlation between $p$ and $\gamma$ is not a meaningful predictor of 3D B-field orientation.  Since the \citet{HalalClark2024} arguments are predicated on this correlation between $p$ and $\gamma$, our results challenge their interpretation that the magnetic field is unlikely to be tangential to the \LB's shell over an appreciable fraction of the surface.

\subsection{Signatures of the Local Bubble in Starlight Polarization Metrics}\label{S:a1_starlight}

We supplement our distance-unresolved \planck-dust polarization observations with distance-resolved starlight polarization observations, assembled by \citet{PanopoulouMarkopoulioti2025} and described in \S\ref{S:starlight_data}.  Various studies of the nearby background starlight polarization measurements have found compelling associations between distance, polarization fraction, and polarization orientation, especially in the context of the \LB \citep[e.g.,][]{Leroy1999, SantosCorradi2011, MedanAndersson2019,GontcharovMosenkov2019, SkalidisPelgrims2019}.

The O24 model of the \LB allows us to directly identify which stars in the \citet{PanopoulouMarkopoulioti2025} catalog fall inside vs. outside of the Bubble's shell.  We expect that LOS that terminate on stars in the \LB's interior should not have polarization properties that correlate with properties of the Bubble's shell (as their light does not pass through the magnetized shell), and that LOS terminating inside vs. outside the shell should consequently have significantly different polarization properties.  We define the shell's inner ($d_{\rm{inner}}$) and outer boundaries ($d_{\rm{outer}}$) using O24's fiducial map at a prominence threshold of $A'_{0.5}$.  Under this definition, 19.5\% of the background stars are in the \LB's interior ($d_\star < d_{\rm{inner}}$), 7.4\% are within the \LB's shell ($d_{\rm{inner}} \leq d_\star \leq d_{\rm{outer}}$), and 73.1\% are located beyond the \LB's shell ($d_\star > d_{\rm{outer}}$). 

We use this information to obtain a distance-resolved view of the effects of the \LB on two quantities: debiased starlight polarization fraction $p_{\rm{d}}$, and starlight polarization angle relative to \planck B-field orientation, 
\begin{equation}
    \Delta \theta_{\rm{planck}} = \frac{1}{n_{\rm{planck}}} \sum_i^{n_{\rm{planck}}} \left| \arctan  \frac{ | \Vec{\Psi}_\perp \times \Vec{\Psi}_\star |}{\Vec{\Psi}_\perp \cdot \Vec{\Psi}_\star} \right| ,
\end{equation}
where agreement is averaged over the $n_{\rm{planck}} = 100$ nearest measurements of \PsiPerp in the \planck map (in 2D projection).  This measure of vector agreement ranges between $\Delta \theta_{\rm{planck}} \in [0^\circ, 90^\circ]$, with \dthetap$=0^\circ$ indicating parallel alignment and \dthetap$=90^\circ$ indicating perpendicular alignment.

\subsubsection{Global Variations in Starlight Polarization Properties}

\begin{figure}
    \centering
    \includegraphics[width=0.48\textwidth]{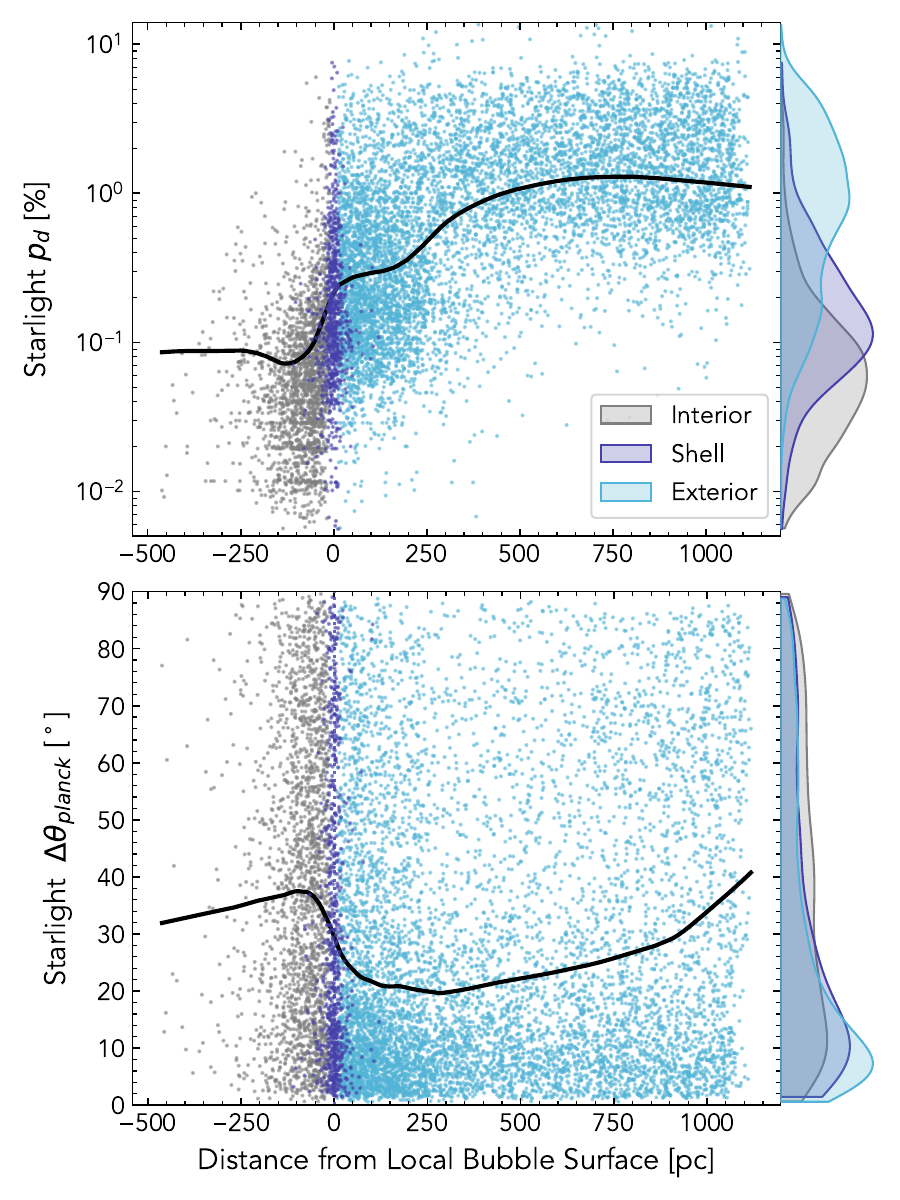}
    \caption{\textit{Top:} Debiased starlight polarization fraction $p_{\rm{d}}$ as a function of stellar distance from the peak extinction surface of the \LB.  Background stars in the \LB's interior are marked in gray, while stars in the shell are shown in purple and stars exterior to the shell are shown in blue.  The black line shows the average relationship between $p_{\rm{d}}$ and distance from the \LB's surface, smoothed with LOWESS regression \citep[Locally Weighted Scatter-plot Smoothing, ][]{cleveland_robust_1979} with an inclusion fraction of 0.2.  On the right edge of the figure, kernel density estimations of $p_{\rm{d}}$ are shown for each stellar distance group.  \textit{Bottom:} As top, but for starlight polarization orientation relative to \planck B-field orientation, $\Delta \theta_{planck}$. }
    \label{fig:global_star_p_theta}
\end{figure}

\begin{figure*}
    \centering
    \href{https://theo-oneill.github.io/magneticlocalbubble/starlight_LOS/}{\includegraphics[width=\textwidth]{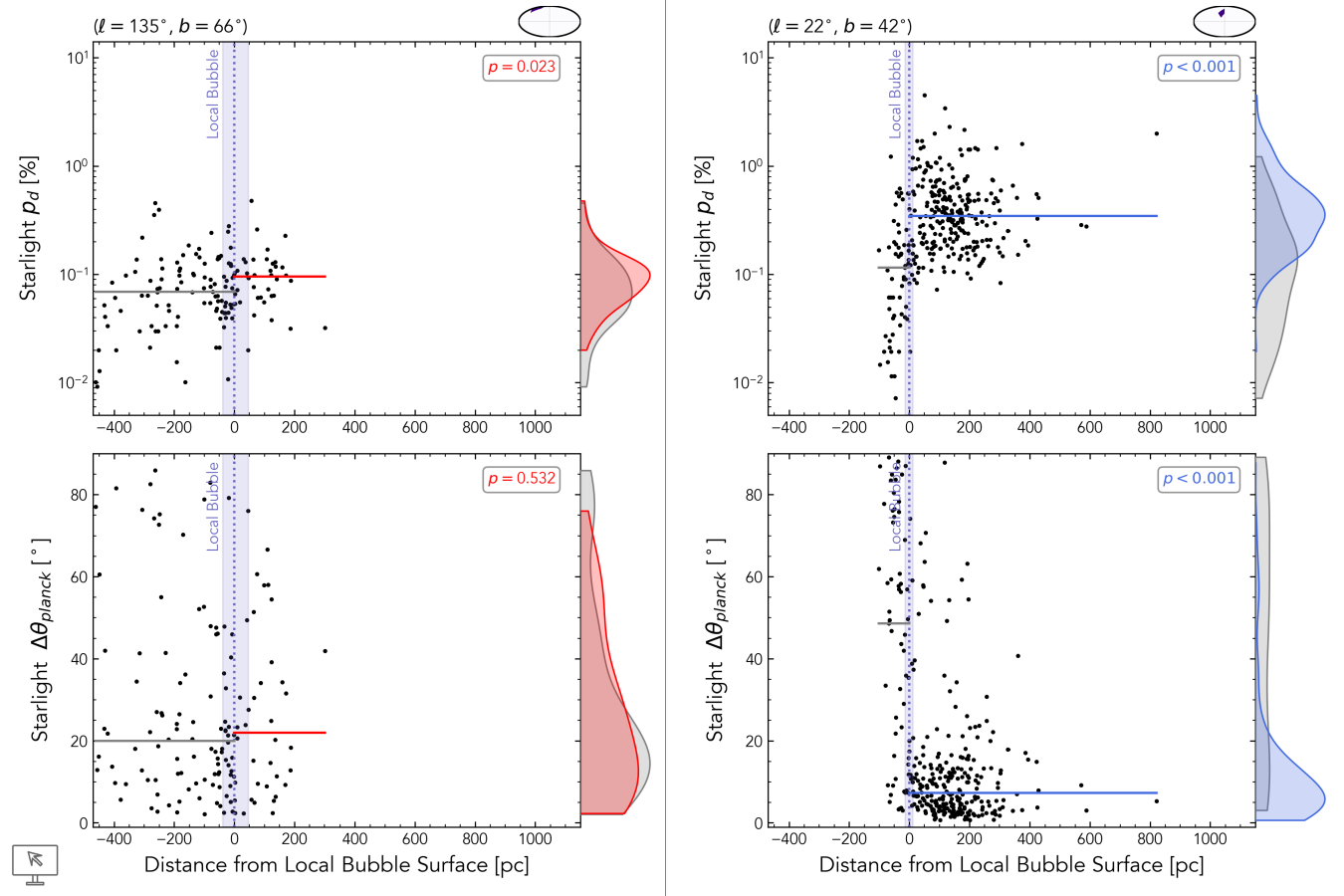}}
    \caption{\textit{Left and right:} As Figure \ref{fig:global_star_p_theta}, but for starlight polarization observations located along two distinct \Hpx \nside{2} cells; the column on the left shows a cell where the \LB's shell has no significant effect on polarization percentage or orientation, while the column on the right shows a cell where both polarization percentage and orientation are significantly different inside vs. outside the \LB's shell.  Central coordinates and on-sky location of the \Hpx cells are shown at the top of each figure.  The average extent of the \LB's inner and outer edge within the cell is marked in purple.  The median values of starlight polarization metrics inside vs. outside the \LB's surface are marked with horizontal lines.  $p$-values for one-sided Kolmogorov Smirnov tests comparing the distribution of internal vs. external properties are reported in inset text boxes.  If $p \geq 0.01$ (as is the case for the \Hpx cell on the left), the median line and KDE for the external starlight sample are colored in red; if $p < 0.01$ (as is the case on the right), they are colored in blue.  A full figure set showing equivalent plots for all \Hpx cells analyzed is available online: \url{https://theo-oneill.github.io/magneticlocalbubble/starlight_LOS/}}
    \label{fig:star_los_figset}
\end{figure*}

\begin{figure*}
    \centering
    \includegraphics[width=\textwidth]{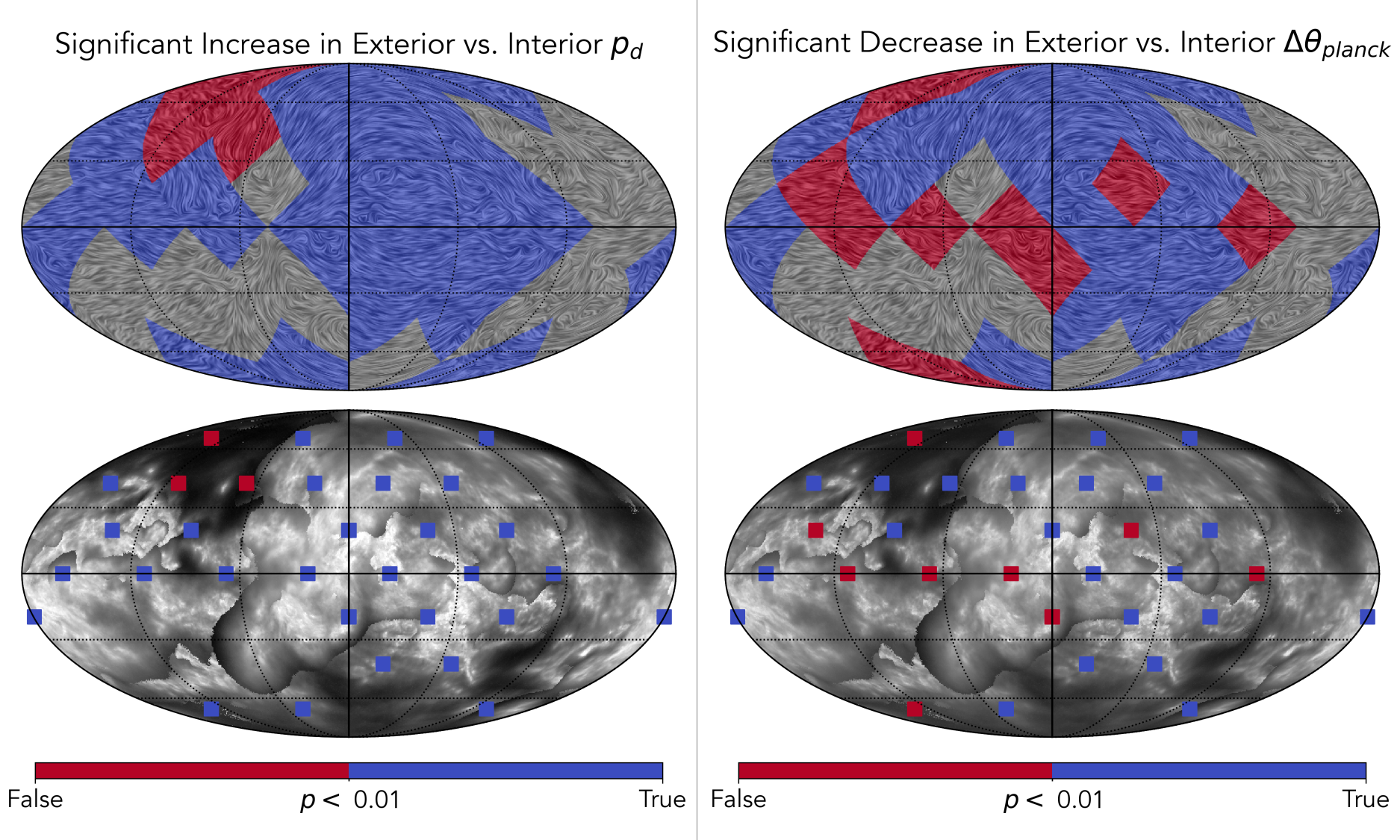}
    \caption{On-sky summary of the results of the starlight polarization significance testing demonstrated in Figure \ref{fig:star_los_figset}.  \textit{Top left:} \Hpx cells with a significant ($p < 0.01$)  increase in $p_{\rm{d}}$ in the exterior starlight group are colored in blue (e.g., the $(\ell, b)=(22\degree, 42\degree)$ cell shown in the right column of Figure \ref{fig:star_los_figset}), while those with no significant increase are colored in red (e.g., the $(\ell, b)=(68\degree, 42\degree)$ cell shown in the left column of Figure \ref{fig:star_los_figset}).  Cells with insufficient sampling of stars are colored in gray.  A Line Integral Convolution of \planck B-field orientation is overlaid.  \textit{Bottom left:} Scatter points marking the central coordinates of analyzed \Hpx cells, with colors as in the top panel, are overlaid on a grayscale map of peak \LB shell density.  \textit{Top right and Bottom right:} As left, but for a significant decrease in starlight polarization angle relative to $\planck$, $\Delta \theta_{planck}$.
    }
    \label{fig:sig_diff_inout}
\end{figure*}

We first consider the global distributions of $p_{\rm{d}}$ and \dthetap in the interior, shell, and exterior of the \LB.  Figure \ref{fig:global_star_p_theta} shows $p_{\rm{d}}$ and \dthetap as functions of distance from the peak extinction surface of the \LB.  Clear separations exist between the distributions of these properties for the interior, shell, and exterior populations.  

The median values of $p_{\rm{d}}$ for stars in the Bubble's interior, shell, and exterior are 0.06\%, 0.14\%, and 0.79\%, respectively.  The median values of \dthetap for stars in the interior, shell, and exterior are 36$^\circ$, 21$^\circ$, and 17$^\circ$.  One-sided Kolmogorov-Smirnov (K-S) tests indicate that stars outside the Bubble have significantly higher $p_{\rm{d}}$ and lower \dthetap than stars inside the Bubble ($p < 0.001$ for both).  \juan{Similarly,} stars in the Bubble's shell have higher $p_{\rm{d}}$ and lower \dthetap than interior stars ($p < 0.001$ for both).  For both properties, the shell population serves as a bridge between the interior and exterior distributions.

These results suggest that the passage of light through the Bubble's magnetized shell has left significant imprints on starlight polarization observations, increasing polarization fraction and changing the orientation of starlight polarization angles towards agreement with \planck.  This latter point suggests that over much of the sample, \planck polarization orientations are set by passage through the Bubble's magnetized shell, in agreement with our \AI.  

We emphasize that significant scatter in both $p_{\rm{d}}$ and \dthetap still exists even in the more homogeneous exterior sample.  Additionally, the overall distribution of \dthetap peaks at a few degrees greater than \dthetap$=0^\circ$, suggesting a slight global offset between the starlight and 353 GHz orientations even in the exterior sample; potential causes of an overall offset between starlight and 353 GHz-derived B-field orientations were discussed in depth by \citet{PlanckCollaborationXII2020}.

\subsubsection{Local Variations in Starlight Polarization Properties}

\begin{figure*}
    \centering
    \includegraphics[width=.75\textwidth]{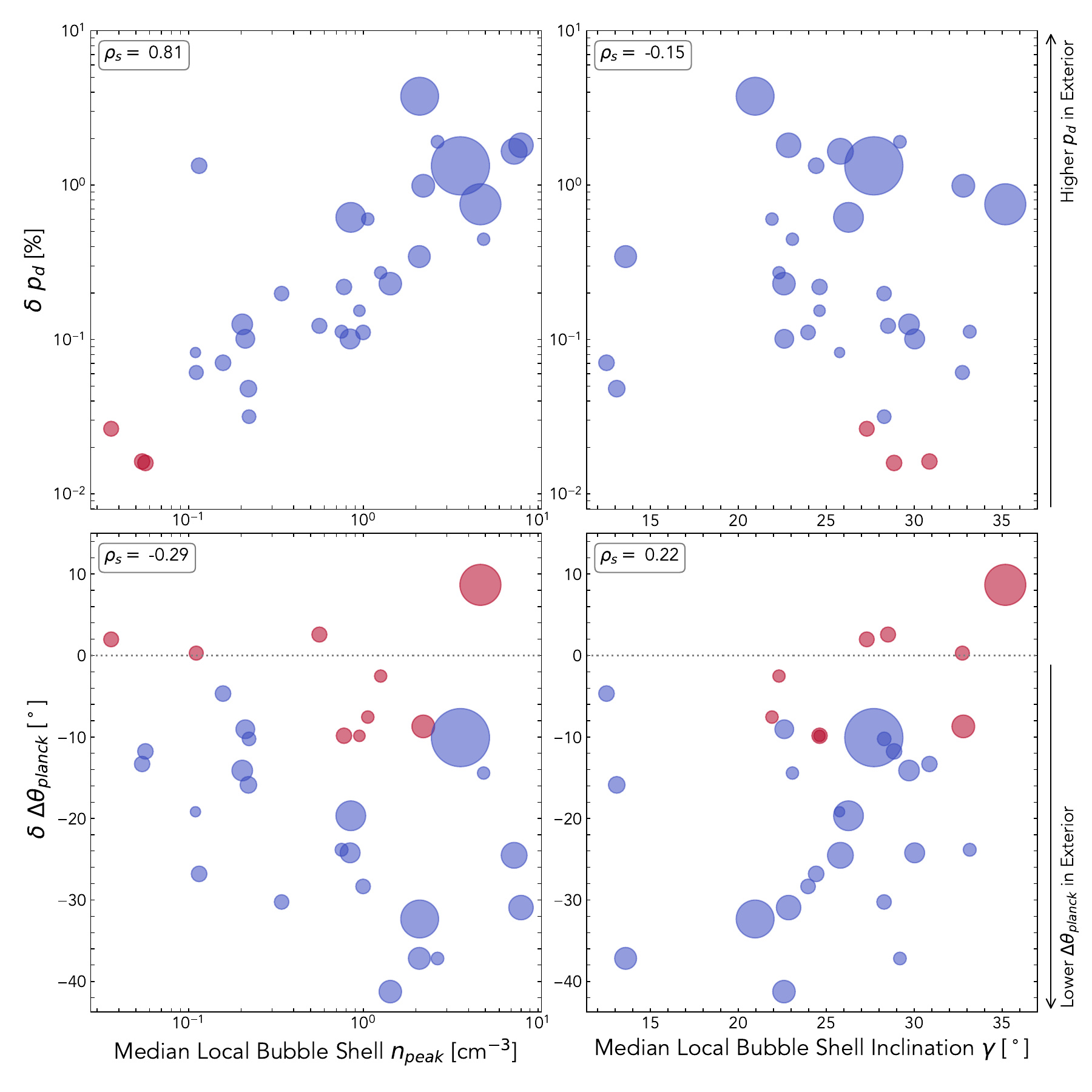}
    \caption{Relationships between properties of the \LB's shell and the difference in median starlight polarization properties inside vs. outside the \LB, for the \Hpx cells analyzed in Figure \ref{fig:star_los_figset}, are displayed.  \textit{Top left:} Change in starlight polarization fraction $\delta p_{\rm{d}}$ vs. median \LB shell density $n_{\rm{peak}}$.  \textit{Top right:} $\delta p_{\rm{d}}$ vs. median \LB shell inclination $\gamma$.  \textit{Bottom left:} Change in starlight polarization orientation relative to \planck $\delta \Delta \theta_{planck}$ vs. $n_{\rm{peak}}$.  \textit{Bottom right:} $\delta \Delta \theta_{planck}$ vs. $\gamma$.  In all figures, points are sized proportionally to the number of stars in the \Hpx cell. \juan{They are} colored according to the results of their K-S tests (with blue marking significant differences).  Spearman correlation coefficients $\rho_s$ are displayed in the inset text boxes.}
    \label{fig:corr_diff_inout}
\end{figure*}

\juan{Significant variations may exist between LOS in the distributions of $p_{\rm{d}}$ and \dthetap, particularly toward LOSs where these properties are the most affected by the Bubble's shell.}  In this section, we examine variations in $p_{\rm{d}}$ and \dthetap in specific regions of the sky.   

To this end, we bin the sky into a \Hpx grid of \nside{2}, yielding 48 equal-area cells (each spanning roughly 29\degree\,$\times$\,29\degree).  To include a cell in our analysis, we require that a minimum of 60 starlight LOS fall within the cell's boundaries, of which at least 30 must terminate on background stars located within the Bubble's interior \juan{and} 30 within the Bubble's exterior.  We verify that our results are reasonably insensitive to the choice of minimum star count.  We emphasize that each cell in this grid covers \juan{a relatively} large area on-sky, and that the coverage of stars within each cell is not uniform. 

With these requirements, 65\% of the sky (31/48 cells) may be included in our experiments.  Figure~\ref{fig:star_los_figset} shows $p_{\rm{d}}$ and \dthetap as a function of distance from \LB's surface for two of these cells; similar plots for all 31 cells analyzed are available online. Analysis of individual LOS reveals that, for the majority of cells, $p_{\rm{d}}$ increases when the surface of the \LB is reached, and that \dthetap transforms from being relatively randomly distributed to clustering around low values (i.e., being parallel to the local \planck B-field orientation).    

For each cell, we perform one-sided K-S tests between the interior and exterior starlight populations to test for significant differences in the distributions of $p_{\rm{d}}$ and \dthetap (with alternative hypotheses chosen to test for an increase in $p_{\rm{d}}$ and decrease in \dthetap in the exterior population).  Figure \ref{fig:sig_diff_inout} shows the results of these tests.  At a significance threshold of $p$-value $< 0.01$, 90.3\% of cells have a significant increase in $p_{\rm{d}}$ outside the \LB, and 71.0\% have a significant decrease in \dthetap. At a more generous threshold of $p < 0.05$, these values would increase to 100\% for $p_{\rm{d}}$ and 77.4\% for \dthetap.  

In the bottom panel of Figure \ref{fig:sig_diff_inout}, we overplot the centers of the considered cells on a map of \juan{the} peak density of the \LB's shell, $n_{\rm{peak}}$.  For $p_{\rm{d}}$, we observe that the three cells with no significant change are located along the low-density Northern ``Local Chimney'' identified by O24.  For \dthetap, one cell with no significant difference is located in the highest-latitude portion of the Local Chimney, in addition to a number of cells with no significant change located at lower latitudes. 

Figure \ref{fig:corr_diff_inout} shows the correlation between the median properties of the \LB's shell in a cell (peak density $n_{\rm{peak}}$ and inclination $\gamma$) and the difference in the median value of $p_{\rm{d}}$ and \dthetap for starlight LOS ending within vs. outside the \LB, which we define as
\begin{equation}
\begin{split}
    \delta p_{\rm{d}} & = \textrm{med}(p_{\rm{d}}[d \geq d_{\rm{LB}}]) - \textrm{med}(p_{\rm{d}}[d < d_{\rm{LB}}]) \\ 
    \delta \Delta \theta_{planck} & = \textrm{med}( \Delta \theta_{p}[d \geq d_{\rm{LB}}]) - \textrm{med}( \Delta \theta_{p}[d < d_{\rm{LB}}]).
\end{split}
\end{equation}
We find a strong positive correlation between $n_{\rm{peak}}$ and $\delta p_{\rm{d}}$ ($\rho_s = 0.81$, $p \ll 0.001$), suggesting that \juan{higher-density} regions of the \LB's shell create a larger imprint on polarization fraction.  We find only insignificant, weak correlations for our other considered trends: $n_{\rm{peak}}$ and $\delta \Delta \theta_{planck}$ ($\rho_s = -0.29$, $p=0.11$), $\gamma$ and $\delta p_{\rm{d}}$ ($\rho_s = -0.15$, $p=0.41$), and $\gamma$ and $\delta \Delta \theta_{planck}$ ($\rho_s = 0.22$, $p=0.23$).

In summary, we find that starlight polarization measurements that sample portions of the LOS outside of the \LB on average have higher polarization fractions and are more aligned with the \planck B-field orientation than starlight polarization measurements that terminate on stars inside the Bubble.  These effects are \juan{susceptible} to position and shell density, with no significant differences originating along the low-density Local Chimney identified by O24.  A decrease in \dthetap occurs even in large portions of the Galactic plane, where the ratio of extinction from the \LB to total extinction along the LOS ($R_{\rm{LB}}$, Figure \ref{fig:ratio_extinc}) is extremely low --- suggesting that even in these extremely confused LOS, \planck polarization orientation may still be dominated by passage through the last significant surface along the LOS, i.e., the \LB, and that our \AI holds.  These conclusions are limited by uneven spatial sampling of starlight polarization measurements inside and outside of the Local Bubble; a more uniform and much \juan{more extensive} catalog of background polarization measurements would be needed to probe these effects at smaller angular scales.

\section{Assumption $\textrm{II}$: Tangency of the Local Bubble's B-field and Shell}\label{S:assum2}

We now consider the second guiding assumption of this work: that the \LB's 3D B-field is tangent to its \juan{dust-traced} shell.  Directly testing this theoretically-predicted result is not possible with current datasets; we must instead turn to simulations of superbubbles expanding into a magnetized ISM to probe this assumption.

To this end, we apply our B-field projection method (\S\ref{S:geometric}) to synthetic polarization maps created by \citet[][hereafter M23]{MaconiSoler2023} for an observer in the interior of a simulated superbubble analogous to the \LB.  We then compare the true orientation of the 3D B-field at the simulated bubble's surface with the projected 3D B-field inferred from our projection method.  We describe our results in full in Appendix \ref{ap:sim}, and refer the interested reader to that section for a complete description of our methods and results.  Here, we highlight the main takeaways of this analysis that may be applicable to our observed \LB.

As reported in Appendix \ref{s:proj_accuracy}, we first test \AII alone by measuring the difference in orientation between the true 3D B-field in the simulated superbubble's shell and the normal vector to the simulated shell.  We find that the true 3D B-field is significantly closer to tangent to the simulated Bubble's surface than a random field, and falls within 40$\degree$ of tangency for 91\% of simulated LOS on the Bubble's surface (where 40$\degree$ is the $+2\sigma$ quantile of the distribution of vector orientation uncertainties for the observed \LB, as derived in Appendix \ref{ap:vec_errors}).  This indicates that \AII is generally well-motivated over the majority of the simulated bubble's surface.  A weak negative correlation exists between tangent orientation and dust density.   

We then test the combination of \AI and \AII by measuring the difference in orientation between the true 3D B-field and the projected 3D B-field.  We observe that the true 3D B-field's orientation is significantly closer to parallel to the projected 3D B-field than a random field, and falls within 40$\degree$ of parallel orientation for 60\% of simulated LOS on the Bubble's surface.  This suggests that the combination of \AI and \AII is well-motivated for the simulated bubble. Weak correlations exist between parallel orientation, increasing B-field strength, and decreasing inclination to the POS.

In summary, Assumptions 1 and 2 appear well-motivated in the context of the simulated bubble.  We conclude that applying these assumptions to the real \LB can likely yield a reasonable model of its true 3D B-field.

\section{Modeling the Local Bubble's 3D Magnetic Field}\label{S:modeling}

Having established that our guiding assumptions are well-motivated, we now describe the results of projecting the 2D \planck polarization observations onto the 3D \LB surface.  We first describe several nuances in visualizing this 3D B-field (\S\ref{S:viz_bfield}), and then report the mean orientation of the B-field in 3D space and in 2D Galactic projection (\S\ref{S:mean_3dB}).  Finally, we model the initial orientation of the local Galactic Magnetic Field in the present-day solar neighborhood in \S\ref{S:gmf_results}.

\subsection{Visualizing the 3D Model}\label{S:viz_bfield}

\begin{figure*}
    \centering
    \includegraphics[width=\textwidth]{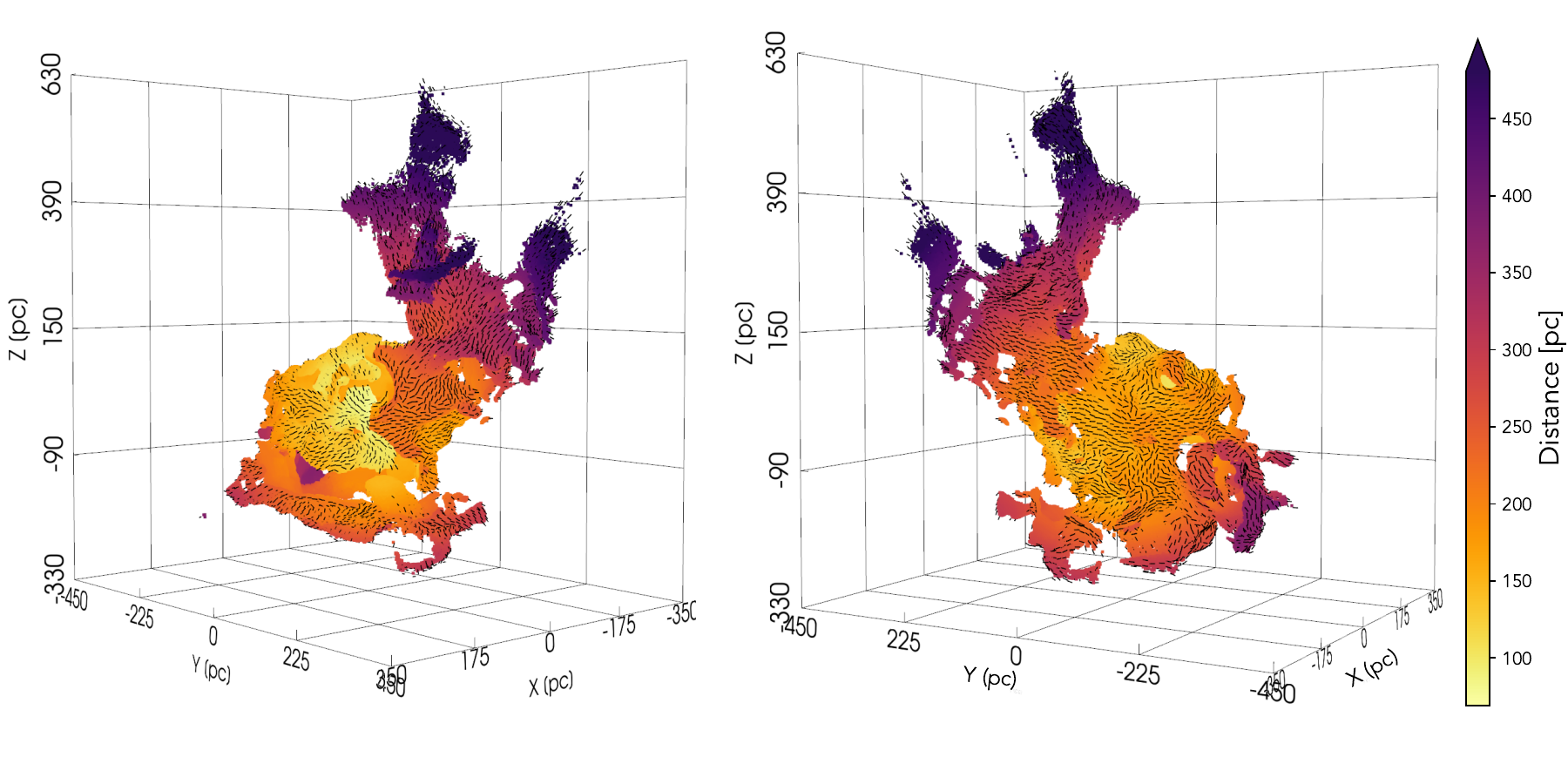}
    \caption{Two views of our 3D model of the \LB's magnetic field.  The 3D B-field model is represented by black vectors overlaid on the \citet{ONeillZucker2024} model of the \LB (colored by distance from the Sun). An interactive figure is available online: \url{https://theo-oneill.github.io/magneticlocalbubble/Bfield/}}
    \label{fig:bfield_3d}
\end{figure*}

Our primary visualization tool for the \LB's 3D B-field is a directionless vector field overlaid on the O24 model of the \LB.  For clarity, we defined a subsample of the full set of $\textbf{B}$ vectors spaced at $\sim$10 pc intervals.  The resulting 10 pc spacing vector field is shown in Figure \ref{fig:bfield_3d}.  The underlying LB model is colored by distance from the Sun to the \LB surface to aid with depth perception.

A machine-readable FITS table of the full \Hpx \nside{256}-spaced representation of the 3D B-field is available online through the Harvard DataVerse.  Since polarization data can only yield pseudovector orientation and not direction, an equivalency $(B_{x}, B_{y}, B_{z}) = (-B_{x}, -B_{y}, -B_{z})$ holds for all 3D vectors.  We also report uncertainties on vector orientation; we discuss our error estimation process in Appendix \ref{ap:vec_errors}.

\subsection{Mean Orientation of the Present-Day 3D B-field}\label{S:mean_3dB}

\begin{deluxetable*}{cc|c|ccc}
\centering
\tablecaption{Mean Orientation of Local Bubble's B-field {\label{tab:summ_bfield}}} 
\tablehead{\colhead{Region} & \colhead{\% LOS} & \colhead{$(\bar{B}_x, \bar{B}_y, \bar{B}_z)$} & \colhead{($\bar{\ell}_{B}$, $\bar{b}_{B}$) [$\degree$]} & \colhead{P20 ($\bar{\ell}_{B}$, $\bar{b}_{B}$) [$\degree$]}  & \colhead{A18 ($\bar{\ell}_{B}$, $\bar{b}_{B}$) [$\degree$]}} 
\startdata
All & \tiny{(100\%)} &(0.474, 0.880, 0.030) & (61.7, 1.7) & & \\
$R_{\rm{LB}} \geq 0.2$ & \tiny{(52.2\%)} & (0.448, 0.893, 0.037) & (63.4, 2.1) & &   \\
Poles & \tiny{(13.4\%)}  & (0.309, 0.951, -0.015) & (72.0, -0.9) & & \\
\hline
North & \tiny{(6.7\%)} & (0.401, 0.887, -0.229) & (65.6, -13.2) & (71.0, -10.9) & (70, 43) \\
South & \tiny{(6.7\%)} & (0.220, 0.968, 0.122) & (77.2, 7.0) & (74.0, 5.8) & (74, -14)
\enddata 
\tablecomments{\% LOS reports the percentage of total lines-of-sight that are included in each region.  Polar regions are defined as $|b| \geq 60\degree$.  ($
\bar{\ell}_B$, $\bar{b}_B$) reports the mean orientation of the B-field in 2D Galactic coordinates, compared to \citet[][P20]{PelgrimsFerriere2020} and \citet[][A18]{AlvesBoulanger2018}.}
\end{deluxetable*}

\citet[][herafter A18]{AlvesBoulanger2018} and \citet[][herafter P20]{PelgrimsFerriere2020} both previously modeled the \LB's magnetic field at high latitudes ($|b| \geq 60\degree$) using \planck 353 GHz polarization maps.  As part of their work, they provided summaries of the present-day orientation of the \LB's B-field in terms of the mean orientation towards Galactic coordinates ($\ell_B$, $b_B$).  Here we present the mean 3D ($\bar{B}_x, \bar{B}_y, \bar{B}_z$) and 2D ($\ell_B$, $b_B$) orientations of our model of the \LB's B-field over various portions of the sky (including the $|b| \geq 60\degree$ polar caps, for direct comparison to the results of A18 and P20). 

We note that, since $\textbf{B} = -\textbf{B}$, simply taking the mean of the full distributions of the 3D B-field's vector components would yield results biased towards the convention adopted for vector direction.  To avoid this effect, we derive the mean 3D orientation $\bar{\textbf{B}} = (\bar{B}_x, \bar{B}_y, \bar{B}_z)$ as the first eigenvector $\textbf{s}_1$ of the scatter matrix, $\bar{\textbf{S}} = \frac{1}{n} \sum_{i=1}^n \textbf{B}_i \textbf{B}_i^T$, a procedure which is appropriate for a Watson distribution of axially-symmetric unit vectors with concentration parameter $\kappa > 0$  \citep{MardiaJupp2000}; see Appendix \ref{ap:vec_errors} for further discussion of the Watson distribution as applied to $\textbf{B}$.

We provide a summary of these mean orientations in Table \ref{tab:summ_bfield}.  A consequence of the lack of vector direction is that each ($\ell_B$, $b_B$) corresponds to an equivalent ($\ell_B + 180\degree$, -$b_B$).  We derive a global average \ellbB{61.7}{1.7} over the full sky, which remains relatively constant for the subset of LOS with $R_{\rm{LB}}\geq 0.2$ (\ellbB{63.4}{2.1}).  In the polar caps, $\ell_B$ tends to take similar values to the whole sky, while $b_B$ obtains a significant vertical components; values for the Northern cap center on \ellbB{65.6}{-13.2}, and for the Southern on \ellbB{77.2}{7.0}.  

As summarized in Table \ref{tab:summ_bfield}, these orientations are in generally good agreement with P20's results for ($\ell_B, b_B$) in the polar caps; for their \lmax=6 model of the \LB, P20 derived \ellbB{71}{-10.9} and \ellbB{74}{5.8} for the North and South, respectively.  The values derived by A18 (\ellbB{70}{43} and \ellbB{74}{-14} for the Northern and Southern caps, respectively) are similar in $\ell_B$ but dissimilar in $b_B$.  We expect these differences are driven both by differences in choice of \LB model and \planck data release and smoothing scale.  

\subsection{Initial Orientation of the Local Galactic Magnetic Field}\label{S:gmf_results}

In addition to measuring the mean present-day orientation of the \LB's B-field, A18 and P20 also modeled the initial orientation of the B-field in the solar neighborhood before disruption by the \LB.  This is possible under \AII and the assumption of pure radial expansion from a point in the interior of the \LB, which together imply that the present-day orientation of the \LB's B-field encodes information about the initial orientation of the interstellar B-field.  In this section, we infer this initial 3D orientation of the local component of the Galactic Magnetic Field (GMF) from our derived 3D model of the \LB's present-day B-field.

\subsubsection{Mathematical Formulation of the Initial Magnetic Field}

The local GMF $\textbf{B}_G$ can be modeled as consisting of a uniform component \Bo and a turbulent component $\textbf{B}_T$ \citep{HildebrandKirby2009,PlanckIntXLIV_2016},
\begin{equation}
    \textbf{B}_G = \textbf{B}_0 + \textbf{B}_T.
\end{equation}
In this work, we focus on modeling \Bo and neglect $\textbf{B}_T$.  We assume that the orientation of \Bo was uniform at all points in 3D space in the solar neighborhood before the \LB's progenitor supernovae occurred,
\begin{equation}
    \textbf{B}_0(x,y,z) = (B_{x, 0}, B_{y, 0}, B_{z, 0}).
\end{equation}
As the \LB expanded from (a) supernova(e) centered at 3D coordinates $c_0$,
\begin{equation}
    c_0 = (x_0, y_0, z_0),
\end{equation}
\Bo would have been swept up into and become tangent to the \LB's shell (\AII).  Using an approximation of the \LB's shell as very thin, A18 formulated the present-day B-field orientation on the \LB's shell as,
\begin{equation}
    \textbf{B}(\textbf{r}) = \frac{r_0}{r} \frac{\partial r_0}{\partial r} \frac{1}{\textbf{n} \cdot \textbf{e}_r } [\textbf{n} \times (\textbf{B}_0 \times \textbf{e}_r)],
\end{equation}
where $r_0$ is the initial position of each particle swept up in the shell, $r$ is the present day position of those particles, and $\textbf{e}_r$ is the radial basis vector originating from \co,
\begin{equation}
    \textbf{e}_r = \left(\frac{x-x_0}{d_0}, \frac{y-y_0}{d_0}, \frac{z-z_0}{d_0} \right),
\end{equation}
where $d_0 = ((x-x_0)^2 + (y-y_0)^2 + (z-z_0)^2)^{1/2}$.  This formulation assumes that the direction of the \LB's expansion from \co has been solely radial (directed along $\textbf{e}_r$). 
A18 and P20 both used this formulation to fit the initial orientation of \Bo in 2D projection, where \Bo can be considered as being oriented towards some coordinate pair ($\ell_0$, $b_0$).  In this work, we extend A18 and P20's 2D fitting to infer the initial 3D orientation of the B-field $\textbf{B}_0$ and expansion center \co.  Since we are only interested in present-day B-field orientation and not strength, we formulate the present-day modeled B-field orientation $\textbf{B}_M$ as a unit vector,  
\begin{equation}
    \textbf{B}_M = \frac{\textbf{n} \times (\textbf{B}_0 \times \textbf{e}_r)}{|\textbf{n} \times (\textbf{B}_0 \times \textbf{e}_r)|}.
    \label{eqn:BM}
\end{equation} 
We note that Equation \ref{eqn:BM} takes the same general form as our model for the observed present-day B-field (Equation \ref{eqn:bfield3d}).  

\subsubsection{Fitting Method}

The orientation of $\textbf{B}_M$ is fully described by 9 parameters:  the three cartesian components of the normal vector $\textbf{n}$ to the \LB's shell, the three components of the original 3D B-field \Bo, and the center of radial expansion \co.  O24 fit a tangent plane defined by the normal vector $\textbf{n}$ to each point on the \LB's surface.  After considering these fit $\textbf{n}$ as givens, we are left with six free parameters $\theta$ to fit to describe \Bo and \co,
\begin{equation}
    \theta = [B_{0x}, B_{0y}, B_{0z}, x_0, y_0, z_0].
\end{equation}
As noted by P20, there is a degeneracy between \Bo and \co such that \co cannot be constrained in the direction parallel to \Bo.  

Our goal is to infer a combination of parameters $\theta$ that yields a modeled $\textbf{B}_M$ consistent with the observed $\textbf{B}$.  We model the likelihood of the observed B-field at any given point on the Bubble's surface $\textbf{B}_i$ given $\theta$, $p(\textbf{B}_i | \theta)$, with a Watson distribution,
\begin{equation}
\begin{split}
    p(\textbf{B}_i | \theta) & = p(\pm \textbf{B}_i | \textbf{ B}_M(\theta), \kappa_i) \\
    &=\left(\frac{\sqrt{\pi} \textrm{erfi}(\sqrt{\kappa_i})}{2\sqrt{\kappa_i}}\right)^{-1} \exp \left( \kappa_i \left(\textbf{B}_M(\theta)^T \textbf{B}_i \right)^2 \right).
\end{split}
\end{equation}
We discuss the Watson distribution and the estimation of the concentration parameter $\kappa$ in Appendix \ref{ap:vec_errors}.  The log-likelihood of all \B given $\theta$ is the sum of the individual log-likelihoods, $\ln p(\textbf{B} | \theta) = \Sigma_{i=1}^n \ln p(\textbf{B}_i | \theta)$.
 
To evaluate the posterior distribution of the probability of $\theta$ given \B, $p(\theta | \textbf{B}) =p(\textbf{B} | \theta) p(\theta) / p(\textbf{B})$, we perform dynamic nested sampling \citep{HigsonHandley2019} as implemented in the python package {\tt dynesty} \citep{Speagle2020}.  We use multiple bounding ellipsoids and uniform sampling.  We adopt uniform priors on \Bo and truncated normal priors on \co, requiring
\begin{equation}
\begin{split}
\textbf{B}_0 &  \begin{cases}
B_{x, 0} \in [-1, 1] \\
B_{y, 0} \in [0, 1] \\
B_{z,0} \in [-1, 1] \\
(B_{x, 0}^2 + B_{y, 0}^2 + B_{z, 0}^2)^{1/2} = 1 \\
\end{cases} \\
c_0 &  \begin{cases}
    x_0 \sim \mathcal{N}(-10, 100) \in [-190, 190] \\
    y_0 \sim \mathcal{N}(2, 118) \in [-280, 310] \\
    z_0 \sim \mathcal{N}(0, 122) \in [-220, 350] 
\end{cases}
\end{split}
\end{equation}
We restrict $B_{y, 0} \in [0, 1]$; relaxing this restriction causes our posteriors to converge on two equal but opposite-signed solutions (($B_{x, 0}, B_{y, 0}, B_{z, 0}$) and ($-B_{x, 0}, -B_{y, 0}, -B_{z, 0}$)), with no significant change to \co.  We additionally place a joint prior on $(B_{x, 0}, B_{y, 0}, B_{z, 0})$ so that $\textbf{B}_0$ is a unit vector.  Our priors on \co are centered on the geometric center of the O24 model of the \LB, with widths determined by the standard deviation of the model's coordinates.  We truncate our volume at the $\pm 3 \sigma$ quantiles of the O24 model's $A_{0.5}$ ``inner edge'' of the \LB's shell to encode our assumption that \co falls in the interior of the present-day volume of the \LB.  We sample points from the \LB's surface at \Hpx \nside{32} resolution for computational efficiency; we verify that estimates of $\theta$ do not change significantly with sampling resolution.

\subsubsection{Modeling Results}

\begin{figure*}
    \centering
    \includegraphics[width=\textwidth]{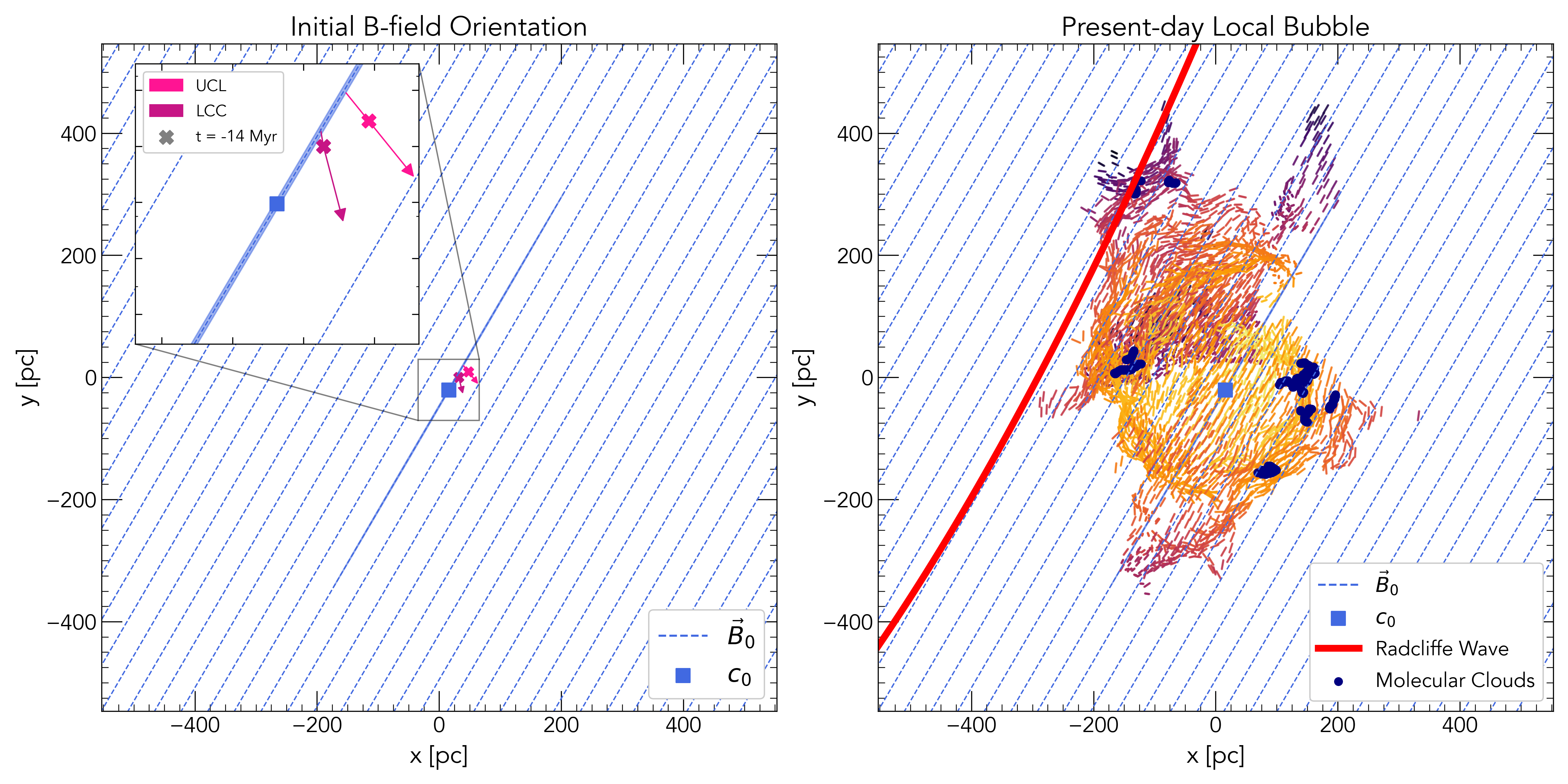}
\caption{Best-fit values for the initial local Galactic Magnetic Field orientation, \Bo, and the \LB's center of radial expansion, $c_0$, are shown in the $x-y$ plane.  \textit{Left:} The orientation of \Bo is shown by the dashed blue vectors extending across the solar neighborhood.  The location of $c_0$ is marked by the blue square, with the 99.7\% (3$\sigma$) credible interval on its location shown by the very thin blue ellipse oriented along \Bo.  The trajectories of the Upper Centaurus Lupus (UCL) and Lower Centaurus Lupus (LCC) stellar clusters derived by \citet{ZuckerGoodman2022} are shown from cluster birth to $t=-10$ Myr by the light pink and dark pink arrows, respectively, with their positions at \citet{ZuckerGoodman2022}'s estimated birth of the \LB at $t=-14$ Myr marked by crosses.  \textit{Right:} As left, but additionally showing a model of the Radcliffe Wave in red \citep{KonietzkaGoodman2024}, the locations of molecular clouds on the \LB's surface in navy \citep{zucker2021}, and this work's 3D model of the \LB's B-field (with vectors colored by distance from the Sun as in Figure \ref{fig:bfield_3d}).}
\label{fig:origBfield_3D}
\end{figure*}

Our posterior distribution of $\theta$ is centered on values of: 
\begin{equation}
\begin{split}
\hat{\textbf{B}}_0 &  \begin{cases}
B_{x, 0} = 0.504 \pm 0.003\\
B_{y, 0} = 0.863 \pm 0.002 \\
B_{z,0}  = -0.027 \pm 0.003 \\
\end{cases} \\
\hat{c}_0 &  \begin{cases}
    x_0 = 15.0 \upm{110.1}{104.9} \ \textrm{pc} \\
    y_0  = -21.8 \upm{188.1}{180.0} \ \textrm{pc} \\
    z_0  = -3.9 \upm{5.7}{6.0} \ \textrm{pc}, 
\end{cases}
\end{split}
\end{equation}
where reported uncertainties mark the 95\% (2$\sigma$) credible intervals.  We emphasize that the uncertainties on $y_0$ and $x_0$ are much larger than on $z_0$ due to the degeneracy between $\textbf{B}_0$ and $c_0$ in the direction of $\textbf{B}_0$.  A corner plot showing the posterior distribution of these variables is presented in Appendix \ref{ap:gmf_model}.     

In Appendix \ref{ap:sim}, we test the accuracy of our B-field orientation inference method using the M23 simulated bubble; we derive an angular distance between our inferred $\hat{\textbf{B}}_0$ and the true initial B-field orientation in the simulated volume of 3.9$\degree$.  This provides a zeroth-order constraint on the accuracy of the initial orientation inferred from the observed \LB in this work.

In 2D Galactic coordinates, our derived $\hat{\textbf{B}}_0$ is oriented towards ($\ell_0$, $b_0$) = (59.70$\degree \pm 0.09\degree$, -1.55$\degree \pm 0.08\degree$).  As discussed, \citet{AlvesBoulanger2018} and \citet{PelgrimsFerriere2020} previously inferred the 2D \Bo from \planck 353 GHz observations at polar ($|b| \geq 60\degree$) latitudes.  Our inferred ($\ell_0$, $b_0$) diverges from their inferred values.  A18 adopted a scale-free elliptical model of the \LB and derived an initial orientation towards ($\ell_0$, $b_0$) = ($71 \pm 11 \degree$, $-16 \pm 7\degree$).  P20 used their 3D dust-derived model of the \LB to infer an initial orientation (for their \lmax = 6 model) towards ($\ell_0$, $b_0$) = ($73.2 \pm 0.1 \degree$, $16.8 \pm 0.4 \degree$) and an expansion center $c_0$ at ($x_0 = 57.6 \pm 34.5$ pc, $y_0 = 79.2 \pm 114.1$ pc, $z_0 = -86.3 \pm 36.1$ pc).  P20 noted that their fit $z_0$ are located significantly below the Galactic plane, which is difficult to reconcile with the expectation of the \LB's progenitor supernovae population being centered in the disk.  We experiment with fitting $\textbf{B}_0$ and $c_0$ based only on the O24 model's polar caps, and derive results in closer agreement to A18 and P20: ($\ell_0$, $b_0$) = ($67.93 \pm 0.46 \degree, 23.57 $ \upm{1.35}{1.43}$\degree$) and an extraplanar $z_0$ = $-57.7$ \upm{91.2}{88.9} pc.  We attribute this result to the lack of geometrical constraints provided by in-disk portions of the \LB's shell, and/or to a potential shift in mean B-field orientation in the present day in the polar caps vs. over the full surface of the \LB.

\citet{KorochkinSemikoz2025} recently modeled the Galactic magnetic field traced by Faraday rotation measures and synchrotron polarization by considering contributions from a variety of large-scale Galactic features, including the Local Bubble.  As part of their analysis, they derived an initial orientation for the B-field in the Solar Neighborhood pre-disruption by the \LB towards ($230.0\degree \pm 1.6\degree$, $-2.0\degree \pm 2.0 \degree$), corresponding to ($\ell_0, b_0$ ) = ($50 \degree, 2\degree$) (given the directional ambiguity in our polarization data) --- in good agreement with our modeled results derived from \planck dust polarization.

\section{Discussion}\label{S:discuss}

Our present-day model of the \LB's B-field is on average oriented towards $\bar{\ell}_B \sim 60 - 70 \degree$ and $\bar{b}_B \sim 0 \degree$, depending on the region of the sky considered.  These values are generally consistent with previous estimates of large-scale interstellar B-field orientations in the Solar Neighborhood.  \citet{PlanckIntXLIV_2016} modeled the structure of the GMF over the Southern polar cap and derived a mean orientation towards \ellbB{70}{24}, which is similar, though not identical, to what we have derived for the Southern cap of the \LB.  Studies of starlight polarization near the Local Bubble have derived orientations towards \ellbB{80}{0} \citep{Heiles1996}, while analysis of pulsar rotation measures \juan{has} yielded $\ell_B = 77\degree$ \citep[][when including LOS that intersect the North Polar Spur]{RandKulkarni1989}. 

Understanding what portion of these measurements arises from the \LB's B-field vs. the more distant, large-scale Galactic B-field requires detailed, 3D studies of polarization as a function of distance and on-sky position.  Through analysis of starlight polarization within 500 pc, \citet{GontcharovMosenkov2019} observed a steep jump in starlight polarization degree at a distance of $\sim$100 pc, which they associated with the transition from the interior to \juan{the} exterior of the \LB.  \citet{SkalidisPelgrims2019} found evidence for the \LB affecting starlight polarization fraction and orientation through analysis of starlight polarization relative to \planck 353 GHz data, while \citet{MedanAndersson2019} analyzed starlight polarization data and demonstrated that the strength of the \LB's magnetic field and the efficiency of grain alignment likely varies significantly across its surface.

Our analysis of dust and starlight polarization metrics in this work demonstrates that passage through the \LB's shell likely has a significant effect on observed polarization over much of the sky, in agreement with previous studies but benefiting from the additional 3D information contributed by the O24 \LB model.  Notably, we find the most consistent and dramatic changes in starlight polarization fraction and orientation relative to \planck occur in the region of the Northern sky above the Galactic center containing the well-known Loop I polarization feature.  Through comparison of starlight polarization as a function of distance relative to synchrotron emission, \citet{Panopoulou2021} derived distance estimates to the Loop I feature ranging between $d < 105$ pc to $d = 135 \pm 20$ pc.  These distance estimates are consistent with the distances to the O24 \LB shell ($91 - 147$ pc) along the same LOS analyzed by \citet{Panopoulou2021}, suggesting that the Loop I polarization orientation originates in the \LB's shell.  

On the whole, our analysis provides support for the idea that a significant fraction of polarization features originate near the Sun at the \LB's magnetized shell.  In addition, we observe a transition from random to ordered starlight polarization orientation as one moves from the interior of the \LB to its shell.  This suggests that at least three components are needed to properly model B-fields observed by \planck and similar facilities: 1) the weakly magnetized interior of the \LB, where polarization orientation is nearly random, 2) the coherent, magnetized shell of the \LB, and 3) the large-scale Galactic magnetic field. 

Finally, we consider what the inferred orientation of the initial Galactic magnetic field (\Bo) in the solar neighborhood can reveal about the evolution of the \LB.  Figure \ref{fig:origBfield_3D} shows the orientation of our best-fit \Bo in the context of the present-day solar neighborhood.  Theory and simulations predict that superbubbles preferentially expand parallel to their environment's initial B-field, and are hindered in expanding in the perpendicular direction \citep[e.g.,][]{Tomisaka1998, NtormousiDawson2017}.  We infer a \Bo that is mostly oriented in the $y$-direction, with a moderate $x$ component and a negligible vertical $z$ component.  The O24 \LB model is more extended in the $y$ and $z$ directions compared to the $x$ direction (with aspect ratios of $\delta y / \delta x = 1.3$, $\delta z / \delta x = 1.4$). Additionally, most of the star-forming molecular clouds on the \LB's surface (Taurus, Chamaeleon, Corona Australis, Musca, Lupus, Ophiuchus, Pipe, \& Cepheus), whose formation \citet{ZuckerGoodman2022} showed was likely triggered by the expansion of the \LB, are situated along the axis perpendicular to \Bo.  

The elongation of the \LB in the direction parallel to \Bo, and its relative compression (and resulting star formation) in the perpendicular direction, is consistent with theoretical predictions for bubble expansion in the magnetized ISM.  We expect that the vertical elongation of the \LB into a northern ``Chimney'' feature is more likely to be the result of vertically-stratified ISM gas density, dynamics, and past star formation activity than of initial B-field strength or orientation; see \citet{ONeillZucker2024} for further discussion.

We additionally consider the relevance of our derived center of radial expansion, $c_0$, in the context of expectations for the feedback-driven evolution of the \LB.  Our best-fit $c_0$ is unconstrained in the direction of \Bo, tracing out a thin region of plausible $c_0$ throughout the interior of the \LB.  By calculating stellar tracebacks, \citet{ZuckerGoodman2022} suggested that the Upper Centaurus Lupus (UCL) and Lower Centaurus Crux (LCC) clusters in the Sco-Cen association may have hosted the progenitor supernovae that drove the \LB's expansion $t=-14$ Myr ago.  We show their derived trajectories of UCL and LCC from cluster birth (at $t=-16$ Myr and $t=-15$ Myr, respectively) until $t= -10$ Myr in Figure \ref{fig:origBfield_3D}.  We find that the UCL and LCC clusters both intersect the range of possible $c_0$ in the $x-y$ plane at their estimated birth times (although they are negatively offset in the $z$ direction by about 15--20 pc).  This general agreement suggests that UCL and LCC are indeed plausible candidates for \juan{launching} the radial expansion of the \LB, as has long been proposed in the literature \citep[see][]{FuchsBreitschwerdt2006,MaizApellaniz2001}.

Finally, we explore the alignment between the \LB's B-field and the local segment of the Radcliffe Wave, a kiloparsecs-long collection of molecular clouds that is likely the gaseous component of the local arm of the galaxy \citep{AlvesZucker2020, SwiggumAlves2022, KonietzkaGoodman2024}.  We observe that the local portion of the Radcliffe Wave is oriented parallel to \Bo in the $x$-$y$ plane.  This alignment is consistent with the theoretical expectation and extragalactic observational evidence that galactic-scale magnetic fields tend to align with spiral structure \citep[see e.g.,][for a review]{BeckWielebinski2013}.  

In total, our analysis suggests that the nascent \LB encountered a magnetic field aligned with \juan{the} Galactic spiral structure, and that the \LB's subsequent expansion and resultant molecular cloud and star formation proceeded to preferentially occur along axes determined by the local magnetic field.  

\section{Conclusions}\label{S:conclude}

In this work, we derive a 3D model of magnetic field orientation on the surface of the Local Bubble by combining \planck dust polarization observations, optical starlight polarimetry, and assumptions about the behavior of magnetic fields on the surface of superbubbles.  Our main conclusions are as follows:

\begin{enumerate}

\item We introduce a novel method to reconstruct 3D B-field orientation from 2D polarization orientation (\S\ref{S:reconstruct}) and apply it to the Local Bubble. Application of this method to a simulated superbubble also indicates it provides a reasonable representation of the true 3D B-field structure (\S\ref{S:assum2}, Appendix \ref{ap:sim}).  

\item Significant correlations exist between properties of the \LB's shell (such as density and inclination to the plane-of-the-sky) and metrics of both dust and starlight polarization (\S\ref{S:assum1}).  Over most lines-of-sight, starlight polarization fraction increases and starlight polarization orientation transitions towards alignment with \planck polarization orientation at the \LB's shell.

\item We model the initial orientation of the Galactic Magnetic Field in the solar neighborhood (\S\ref{S:modeling}), and find an orientation parallel to the gaseous component of the local arm of the galaxy (the Radcliffe Wave).  We find that the \LB is more extended in the direction parallel to our inferred initial orientation, and that molecular clouds on the \LB's shell are preferentially concentrated in the perpendicular direction.

\end{enumerate}

Overall, this work suggests that the \LB's magnetized shell significantly influences our view of Galactic polarization, and that the \LB's evolution has been shaped by interstellar magnetic fields.  Future efforts to reconstruct the 3D B-field on the \LB's surface --- independent of our \AI and \AII --- could combine starlight polarization tomography \citep[e.g.,][]{PanopoulouTassis2019,PelgrimsPanopoulou2023,PelgrimsMandarakas2024} with line-of-sight B-field probes such as Faraday rotation measures \citep[e.g.,][]{Ferriere2016,TahaniPlume2018,TahaniGlover2022}.  Upcoming \juan{extensive} polarization surveys such as PASIPHAE \citep{TassisRamaprakash2018} will play a crucial role in \juan{the construction of} higher resolution, tomographic decompositions of the \LB's magnetic field.

Finally, corrections for Galactic polarization in studies of the Cosmic Infrared/Microwave Backgrounds \juan{and modeling of the Galactic scale magnetic fields} should account for local foreground contributions by the \LB.  \juan{Identifying the structure of the \LB's magnetized shell offers a path toward correcting the foreground contribution to these observations and Galactic-scale models.
Furthermore, it will reveal in detail the interplay between magnetization and stellar feedback in the nearest laboratory of ISM physics.}

\newpage
\begin{acknowledgements}
We thank Efrem Maconi, Susan Clark, George Halal, Ralf Klessen, Gina Panopoulou, Jo\~{a}o Alves, and Micah Acinapura for insightful discussions.  T.J.O. acknowledges that this material is based upon work supported by the National Science Foundation Graduate Research Fellowship under Grant No. DGE 2140743.  T.J.O., A.A.G., and C.Z. acknowledge support by NASA ADAP grant 80NSSC21K0634 ``Knitting Together the Milky Way: An Integrated Model of the Galaxy’s Stars, Gas, and Dust.''   T.J.O. and A.A.G. acknowledge support from National Science Foundation grants 2209623 and 1908419.  The SAO REU program is funded in part by the National Science Foundation REU and Department of Defense ASSURE programs under NSF Grant no. AST-2050813, and by the Smithsonian Institution (T.J.O.).  J.D.S is funded by the European Research Council via the ERC Synergy Grant ``ECOGAL -- Understanding our Galactic ecosystem: From the disk of the Milky Way to the formation sites of stars and planets'' (Project ID 855130).  The authors acknowledge Interstellar Institute's program "With Two Eyes" and the Paris-Saclay University's Institut Pascal for hosting discussions that nourished the development of the ideas behind this work. 
\end{acknowledgements}

\vspace{1em}
{\large \textit{Interactive Figures:}} Interactive figures presented in this work can be accessed at: \url{https://theo-oneill.github.io/magneticlocalbubble/}.  

\vspace{1em}
{\large \textit{Data Access:}} The derived models of the Local Bubble's 3D Magnetic Field can be downloaded from the Harvard Dataverse: \url{https://doi.org/10.7910/DVN/A8HWUF}, including:
\begin{itemize}
    \item Table of \LB's 3D B-field orientation
    \item Standalone HTML files of interactive figures
\end{itemize}

\software{Astropy \citep{astropy_2013,astropy_2018,AstropyCollaborationPriceWhelan2022};  Cmasher \citep{cmasher2020}; connected-components-3d \citep{connectedcomponents3d}; Dustmaps  \citep{M_Green_2018}; dynesty \citep{Speagle2020}; glue \citep{RobitailleBeaumont2019}; Healpy \citep{Zonca2019_healpy}; K3d-jupyter \citep{k3d_jupyter}; Matplotlib \citep{matplotlib_Hunter2007}; Numpy \citep{harris2020_numpy}; OpenSpace \citep{Bock2017_openspace}; Pandas \citep{pandas_mckinney-proc-scipy-2010}; PyVista \citep{sullivan2019pyvista}}

\appendix
\restartappendixnumbering

\section{Uncertainty Estimation on 3D B-field}\label{ap:vec_errors}
\restartappendixnumbering

\begin{figure*}
    \centering
    \includegraphics[width=\textwidth]{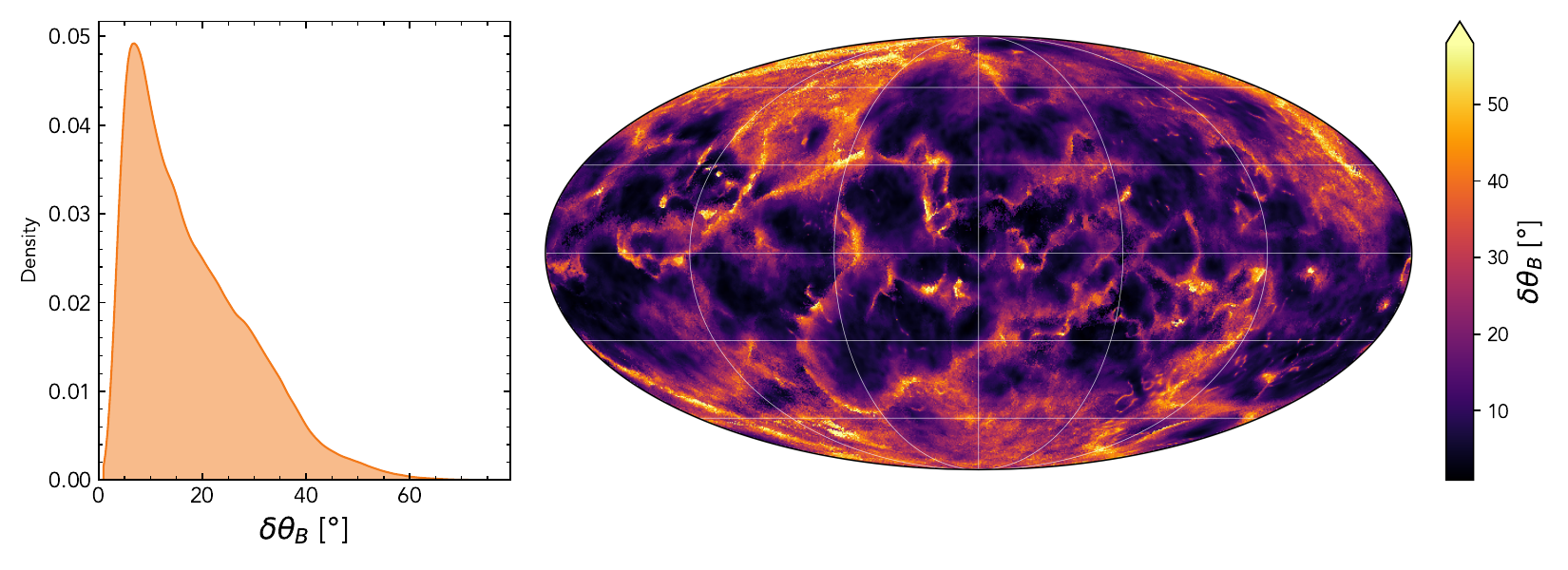}
    \caption{\textit{Left:} Kernel density estimation of average uncertainty on 3D vector orientation, $\delta \theta_B$, derived from our Monte Carlo experiments.  \textit{Right:} Mollweide projection of $\delta \theta_B$.}
    \label{fig:delta_theta_B}
\end{figure*}

Our derived 3D magnetic field model \B is a field of axially symmetric unit vectors, i.e., vectors defined by orientation but not direction ($\textbf{B} = -\textbf{B}$), with associated uncertainties.  A natural model for this field is the Watson distribution \citep{Watson1965equatorial,MardiaJupp2000}, which describes axially symmetric $p$-dimensional vectors on the $S^{p-1}$ unit hypersphere.  The Watson distribution's probability distribution function for $p=3$-dimensional vectors is given by,
\begin{equation}
   p(\pm \textbf{B} | \boldsymbol{\mu}_B, \kappa_B) = \mathcal{M}\left(\frac{1}{2}, \frac{3}{2}, \kappa_B\right)^{-1} \exp \left( \kappa_B (\boldsymbol{\mu}_B^T \textbf{B})^2 \right),
\end{equation}
where $\textbf{B} = (B_{x}, B_{y}, B_{z})^T$, $\boldsymbol{\mu}_B = (\mu_{B,x}, \mu_{B,y}, \mu_{B,z})^T$ is the mean orientation, and $\mathcal{M}\left(\frac{1}{2}, \frac{p}{2}, \kappa\right)$ is the Kummer function,
\begin{equation}
    \mathcal{M}\left(\frac{1}{2}, \frac{3}{2}, \kappa_B\right) = \frac{\sqrt{\pi} \ \textrm{erfi}(\sqrt{\kappa_B})}{2\sqrt{\kappa_B}}.
\end{equation}
$\kappa_B$ is a parameter describing the concentration of the probability distribution function around $\pm \mu_B$; larger positive values of $\kappa_B$ correspond to greater concentration.  Note that this distribution enforces rotational symmetry around the axis defined by $\pm \mu_B$.  

To quantify our uncertainties on \B=\B(\PsiPerp, $\textbf{n}$), we must estimate the parameters of the Watson distribution $\mathcal{W}(\boldsymbol{\mu}_B, \kappa_B)$ describing each B-field vector.  We estimate these parameters via a Monte Carlo approach, where we incorporate two main sources of uncertainty: uncertainties contributed by the \planck dust polarization angles \PsiPerp, and uncertainties contributed by the orientation of the \LB's shell in 3D space as defined by the normal vector $\textbf{n}$.   

\subsection{Planck-based Uncertainties}

To estimate \planck-derived uncertainties, we first generate versions of the \planck GNILC $I$, $Q$, and $U$ covariance maps smoothed to our adopted $2\degree$ resolution.  Following \citet{PlanckCollaborationXII2020}, we consider these observed parameters as originating from a multivariate Gaussian distribution centered at the \planck measurements $\boldsymbol{\mu}_{planck} = (I, Q, U)^T$ with covariances defined as $\mathbb{C}_{planck}$.  Since \PsiPerp is defined by $Q$ and $U$, we can then generate random realizations of \PsiPerp by sampling from $(I, Q, U) \sim \mathcal{N}(\boldsymbol{\mu}_{planck}, \mathbb{C}_{planck})$.

\subsection{Inclination-based Uncertainties}

To estimate inclination-derived uncertainties, we make use of the $n_{\rm{draw}} = 12$ draws of the \citet{EdenhoferZucker2024} dust map, which O24 \juan{used} to characterize uncertainty in their model of the \LB.  As part of this analysis, O24 derived a model of the \LB from each draw and fit all associated properties including $\textbf{n}$.  Since $\textbf{n}$ is a directed unit vector, we can consider its distribution as being characterized by a 3D von Mises-Fisher (vMF) distribution \citep{VonMises1918,Fisher1953,MardiaJupp2000}, which describes directed $p$-dimensional vectors on the $S^{p-1}$ unit hypersphere.  The vMF distribution's probability distribution function for $p=3$ is given by,
\begin{equation}
   p(\textbf{n} | \boldsymbol{\mu}_n, \kappa_n) = \left(\frac{\kappa_n}{4\pi \sinh \kappa_n}\right) \exp \left( \kappa_n (\boldsymbol{\mu}_n^T \textbf{n}) \right),
\end{equation}
where $\textbf{n} = (n_{x}, n_{y}, n_{z})^T$, $\boldsymbol{\mu}_n = (\mu_{n,x}, \mu_{n,y}, \mu_{n,z})^T$ is the mean direction, and $\kappa_n$ is the concentration parameter (with similar interpretation to $\kappa$ in the Watson distribution).  The mean direction $\boldsymbol{\mu}_n$ can be estimated as
\begin{equation}
    \hat{\boldsymbol{\mu}}_n = \frac{1}{\bar{r}} \left[ \frac{1}{n_{\rm{draw}}} \sum_{j=1}^{n_{\rm{draw}}} \textbf{n}_{\rm{draw}, j} \right],
\end{equation}
where $\bar{r} = \frac{1}{n_{\rm{draw}}}|| \sum_j^{n_{\rm{draw}}} \textbf{n}_{\rm{draw},j} ||$ is the mean resultant length.  Concentration $\kappa_n$ can be estimated from $\bar{r}$ using the approximation derived by \citet{BanerjeeDhillon2005},
\begin{equation}
\hat{\kappa}_n \simeq \frac{\bar{r} (3 - \bar{r}^2)}{1 - \bar{r}^2}.
\end{equation}
We are then free to generate random realizations of $\textbf{n}$ as $\textbf{n} \sim \mathcal{VMF}(\hat{\boldsymbol{\mu}}_n, \hat{\kappa}_{n})$.  

\subsection{Combined Uncertainties}

For each point on the \LB's surface, we produce $n_{\rm{samp}} = 10,000$ random samples from $\mathcal{N}(\boldsymbol{\mu}_{planck}, \mathbb{C}_{planck})$ and $\mathcal{VMF}(\hat{\boldsymbol{\mu}}_n, \hat{\kappa}_{n})$.  We project the resulting $n_{\rm{samp}}$ pairs of $\psi_\perp$ and $\textbf{n}$ to $n_{\rm{samp}}$ 3D B-field vectors $\textbf{B}_{\rm{samp}}$ using the procedure outlined in \S\ref{S:geometric}.  We can now proceed to use $\textbf{B}_{\rm{samp}}$ to characterize $\mathcal{W}(\boldsymbol{\mu}_B, \kappa_B)$.  

The mean direction $\boldsymbol{\mu}_B$ can be estimated as the first eigenvector of the scatter matrix, \citep{MardiaJupp2000},
\begin{equation}
    \textbf{S} = \frac{1}{n_{\rm{samp}}} \sum_{j=1}^{n_{\rm{samp}}} \textbf{B}_{\rm{samp},j} \textbf{B}_{\rm{samp},j}^T,
\end{equation},
\begin{equation}
    \hat{\boldsymbol{\mu}}_B = \textbf{s}_1 \ \textrm{if} \ \hat{\kappa}_B > 0.
\end{equation}
To find the maximum likelihood estimate of $\kappa_B$, we must find the solution to the relationship,
\begin{equation}
    \hat{\boldsymbol{\mu}}_B^T \textbf{S} \hat{\boldsymbol{\mu}}_B =  \frac{\mathcal{M}'\left(\frac{1}{2}, \frac{3}{2}, \hat{\kappa}_B\right)}{\mathcal{M}\left(\frac{1}{2}, \frac{3}{2}, \hat{\kappa}_B\right)}. 
    \label{eqn:hat_kappa_B}
\end{equation}
We solve for $\hat{\kappa}_B$ numerically, as no closed-form solution to Eqn. \ref{eqn:hat_kappa_B} exists.  We impose a maximum $\hat{\kappa}_B = 600$ to prevent overflow in the evaluation of the Watson PDF for very concentrated distributions; 0.2\% of vectors in \B require this upper limit.  Derived values of $\hat{\kappa}_B$ range between 0.40 and 600, with a mean of 32.6.

We additionally summarize the uncertainty on B-field orientation at each point on the \LB's surface by computing the mean angular difference between $\textbf{B}_i$ and $\textbf{B}_{samp}$,
\begin{equation}
    \delta \theta_{{B_i}} =\frac{1}{n_{\rm{samp}}} \sum_{j=1}^{n_{\rm{samp}}} \cos^{-1} (|\textbf{B}_i \cdot \textbf{B}_{\rm{samp}, j}|).
\end{equation}
We provide a histogram and 2D projected view of $\delta \theta_B$ in Figure \ref{fig:delta_theta_B}.  $\delta \theta_B$ ranges between 0.94$\degree$ and 79.4$\degree$, with a mean of 17.8$\degree$.  $\theta_B$ is strongly negatively correlated with \LB shell density $n_{\rm{peak}}$ ($\rho_s = -0.72$, $p < 0.001$), i.e., uncertainties on 3D B-field orientation tend to be lower in high-density regions of the shell, where variation between draws of the \citet{EdenhoferZucker2024} dust map is low.

\section{Performance for a Simulated Local Bubble Equivalent}\label{ap:sim}
\restartappendixnumbering

To test the accuracy of our B-field projection method and the assumptions guiding this work, we apply our method to a simulated analog of the \LB analyzed by \citet[][hereafter M23]{MaconiSoler2023}.  M23 analyzed snapshots from the \citet{GirichidisSeifried2018} and \citet{Girichidis2021} simulations \citep[created as part of the SILCC project, ][]{WalchGirichidis2015,GirichidisWalch2016} of supernova-driven superbubbles evolving in a magnetized (500 pc)$^{3}$ volume of the ISM and identified a superbubble with similar physical properties to the real \LB (including B-field strength and shell density).  Similar to the real \LB, the M23 simulated bubble has also formed an asymmetric Chimney out of the disk.

M23 then generated synthetic 353 GHz polarization maps (Stokes $I$, $Q$, and $U$) for an observer in the center of the bubble using the {\tt POLARIS} radiative transfer code \citep{ReisslWolf2016}.  In this appendix, we combine these simulated 2D polarization maps with the ground-truth 3D B-field orientation within the shell of the simulated bubble to test the accuracy of our 3D B-field projection method, and our assumption of 3D B-field tangency to the superbubble's shell.  M. Kunold et al. (in preparation) will analyze the performance of the assumption of B-field tangency for the simulated bubble in greater depth.

\subsection{Modeling the Simulated Bubble}

\begin{figure*}
    \centering
    \href{https://theo-oneill.github.io/magneticlocalbubble/sim_Bfield/}{\includegraphics[width=\textwidth]{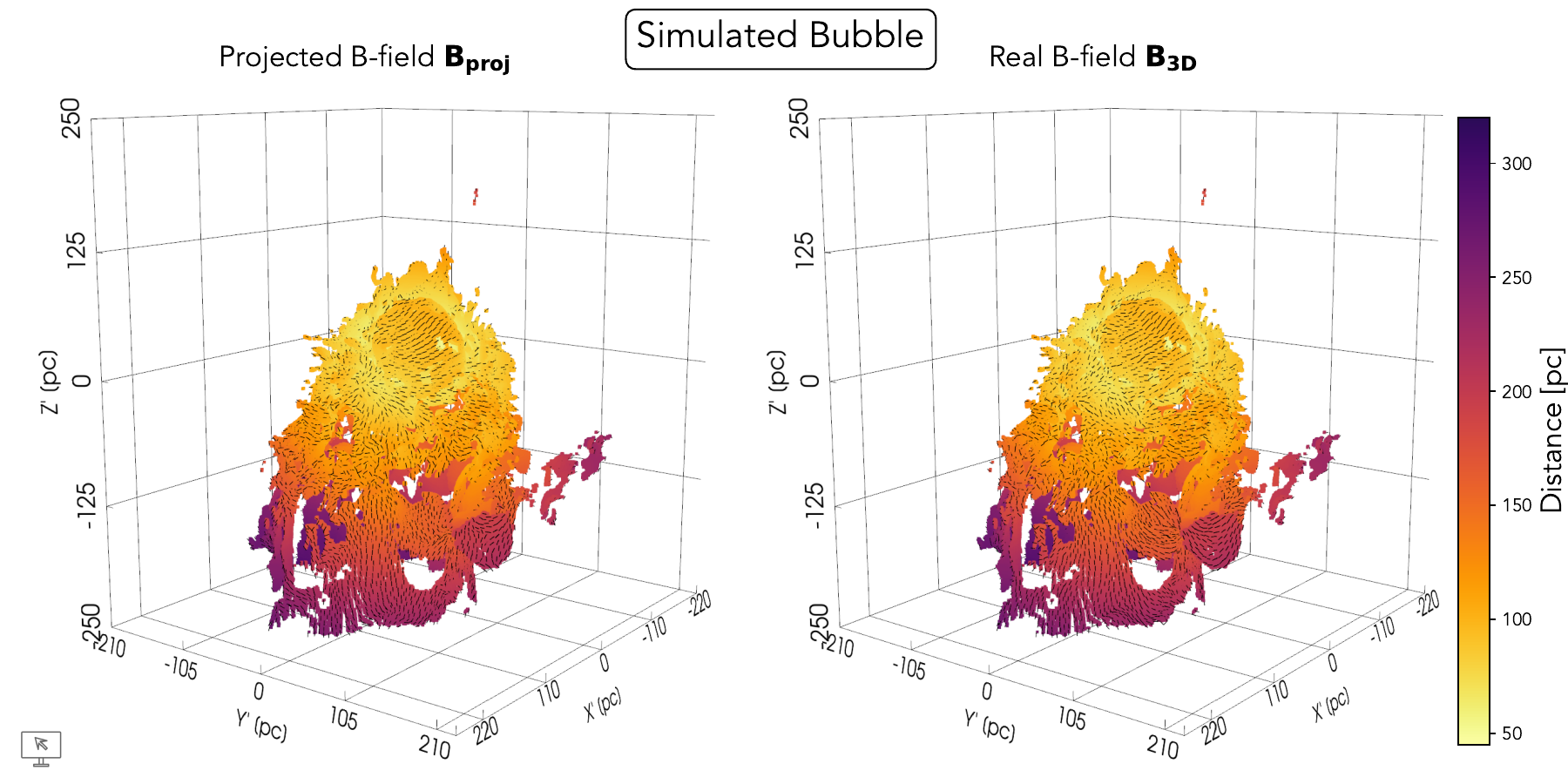}}
    \caption{\textit{Left:} 3D B-field model derived via our fiducial projection method, $\textbf{B}_{\rm{proj}}$, for a simulated equivalent of the \LB.  $\textbf{B}_{\rm{proj}}$ is shown by the black vectors, overlaid on a model of the simulated bubble's shell colored by distance from the observer.  \textit{Right:} As left, but showing the real 3D B-field on the surface of the simulated bubble, $\textbf{B}_{\rm{3D}}$.  Interactive figure is available online at: \url{https://theo-oneill.github.io/magneticlocalbubble/sim_Bfield/}}.
    \label{fig:simbub_map}
\end{figure*}

To assess our method's performance, we first generate a model of the simulated bubble's peak density surface using the same peak-finding method employed by \citet{ONeillZucker2024} to model the \LB's shell.  This method identifies the first prominent peak along the LOS from an observer at the center of the bubble being mapped.  Although this basic principle holds regardless of use case, a key difference between the application of this method to real vs. simulated bubbles is that simulated volumes (like the simulated 3D gas density map analyzed by M23) are able to reveal structures of all densities in and around bubbles, while current 3D dust mapping techniques are only sensitive to a limited range of densities.  Inspection of the M23 gas density map reveals a variety of extremely diffuse filaments and sheets in the interior of the simulated bubble that are not a part of the bubble's shell and would not be resolved by 3D dust maps, but would interfere with our peak-finding procedure.  

To map the simulated bubble's shell using the O24 method while overlooking these ultra-low-density structures, we perform a multi-step cleaning procedure which we emphasize was developed on an ad-hoc basis for the specific conditions of this simulation.  Specific parameters described below were selected via experimentation with an eye towards reducing contamination of the final bubble shell model by interior structures.  Working in cartesian space, we first identify and mask all material in the gas cube below a density of  $n = 10^{-1}$ cm$^{-3}$.  This removes the majority of low-density contaminants but leaves behind several isolated intermediate-density cores in the bubble's interior.  We identify these isolated cores in our thresholded map using the python package {\tt connected-components-3D} \citep{connectedcomponents3d} and mask them out as well.  We additionally mask pixels with low B-field strengths ($|B| < 1 \mu$G).  For gas more than 3 scale-heights ($H = 30$ pc, \citealt{GirichidisSeifried2018}) above and below the simulated bubble's center, we perform a similar procedure with a gas density threshold of $n= 10^{-2}$ cm$^{-3}$ and no B-field strength requirement.  In total, this cleaning process yields a cartesian gas density map analogous to what might be resolved by current 3D dust mapping techniques and suitable for application of the O24 peak-finding method.

We project this cleaned gas density map to a spherical coordinate system defined at \Hpx \nside{256} sampling with 1 pixel (0.98 pc) radial spacing.  We apply the O24 peak-finding method to this spherical volume, smoothing each LOS with a Gaussian kernel of $\sigma = 3$ pixel (2.94 pc) and requiring a minimum prominence of $10^{-2}$ cm$^{-3}$.  This yields a 3D map of the simulated bubble with the same set of properties derived by O24 in their model of the \LB, including shell distance, density, and inclination.  We then apply the B-field projection procedure described in \S\ref{S:geometric} to the simulated bubble.  We first match the M23 simulated 2D Stokes $I$, $Q$, and $U$ polarization maps to each LOS in the simulated Bubble's shell, and then project the corresponding B-field angles to 3D to infer the B-field orientation $\textbf{B}_{\rm{proj}}$.  The resulting 3D models of the bubble and $\textbf{B}_{\rm{proj}}$ are shown in Figure \ref{fig:simbub_map}.

\subsection{Accuracy of Projection Method}\label{s:proj_accuracy}

\begin{figure}
    \centering
    \includegraphics[width=.48\textwidth]
    {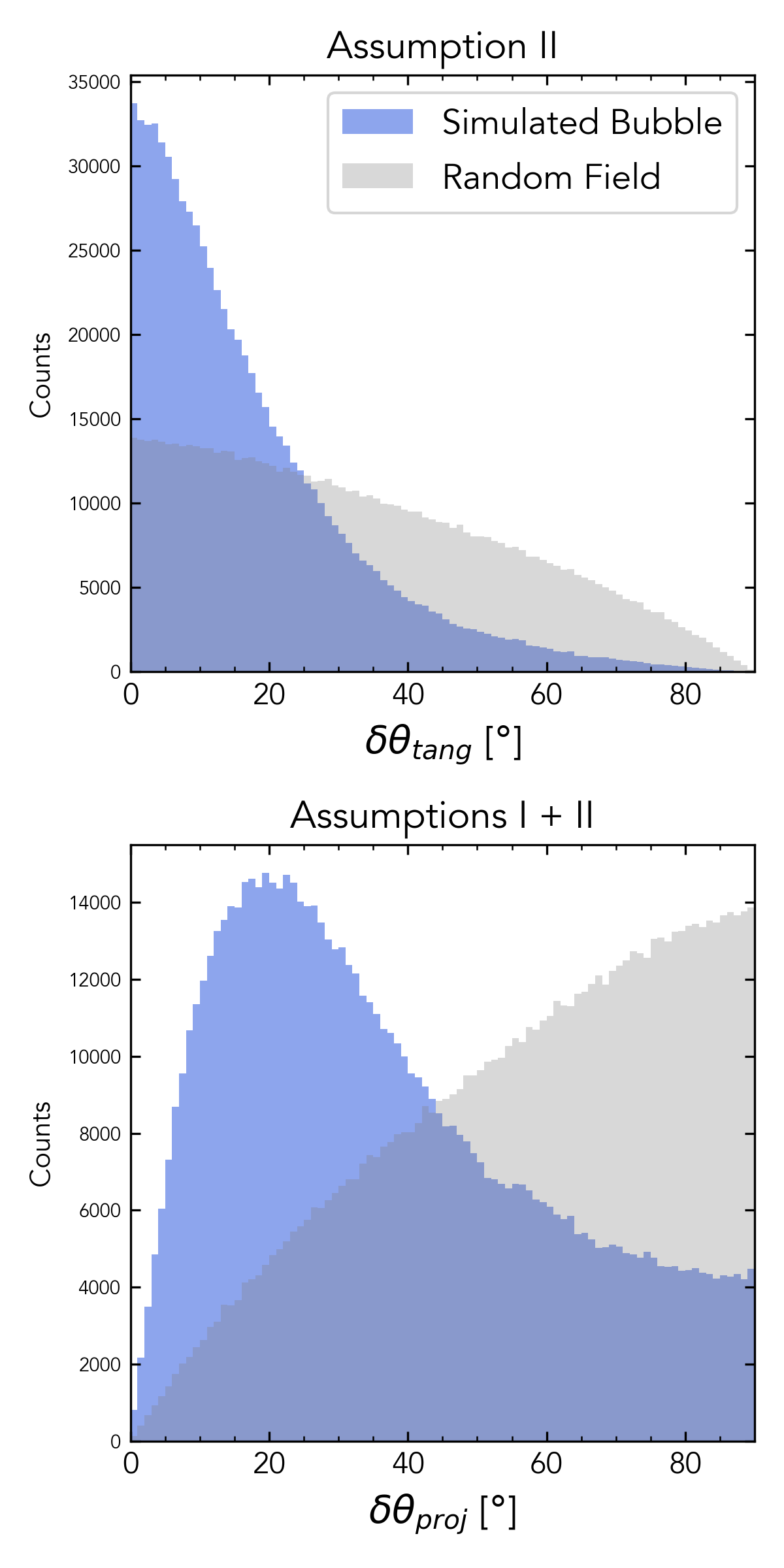}
    \caption{\textit{Top:} Histogram of the performance of our \AII for the simulated bubble, as traced by our defined metric $\delta \theta_{\rm{tang}}$, is shown in blue.  The performance of this metric for a randomly oriented field is shown in gray.  \textit{Bottom:} As top, but for the performance of our combined \AI and \AII as traced by a metric $\delta \theta_{\rm{proj}}$.}
    \label{fig:sim_tang_proj}
\end{figure}

\begin{figure}
    \centering
    \includegraphics[width=0.48\textwidth]{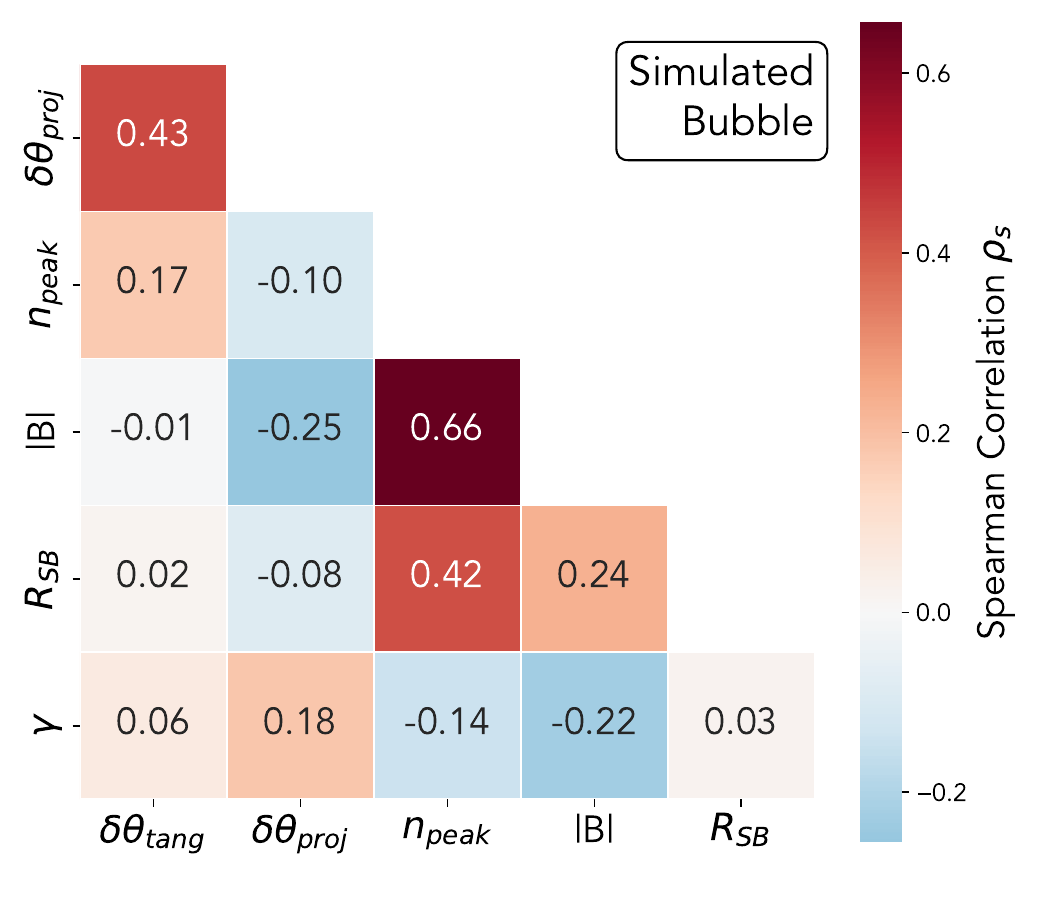}
    \caption{Matrix of Spearman correlation coefficients between $\delta \theta_{\rm{tang}}$, $\delta \theta_{\rm{proj}}$, and properties of the simulated bubble's shell ($n_{\rm{peak}}$, $|B|$, $R_{\rm{SB}}$, and $\gamma$).}
    \label{fig:sim_corr}
\end{figure}

Our B-field projection model relies on the assumptions that: I) polarization orientation originates at the bubble's magnetized shell, and II) the bubble's B-field is tangent to the bubble shell.  We can directly test the validity of these assumptions for the M23 simulated bubble by comparing our projected B-field orientation $\textbf{B}_{\rm{proj}}$ to the true orientation of the B-field at each point on the shell's surface $\textbf{B}_{\rm{3D}}$.  $\textbf{B}_{\rm{proj}}$ and $\textbf{B}_{\rm{3D}}$ are both shown in Figure \ref{fig:simbub_map}.  

We first quantify the validity of our tangency assumption (\AII) by calculating the difference in orientation between $\textbf{B}_{\rm{3D}}$ and the tangent plane fit to the simulated bubble's surface (characterized by the normal vector $\textbf{n}$).  We define a metric
\begin{equation}
    \delta \theta_{\rm{tang}} = 90\degree - \cos^{-1}(|\textbf{B}_{\rm{3D}} \cdot \textbf{n}|),
\end{equation}
where $\delta \theta_{\rm{tang}} = 0\degree$ indicates parallel alignment and $\delta \theta_{\rm{tang}}=90\degree$ indicates perpendicular alignment.  Figure \ref{fig:sim_tang_proj} shows a histogram of $\delta \theta_{\rm{tang}}$ compared to the same statistic derived for two random 3D B-fields.  $\delta \theta_{\rm{tang}}$ has a median value of 13$\degree$.  As assessed by a one-sided Kolmogorov-Smirnov test, the distribution of $\delta \theta_{\rm{tang}}$ takes on values significantly closer to parallel alignment than for a random field ($p < 0.001$).  

We then quantify the combined validity of \AI and \AII by calculating the agreement between $\textbf{B}_{\rm{3D}}$ and $\textbf{B}_{\rm{proj}}$, defining a similar metric
\begin{equation}
    \delta \theta_{\rm{proj}} = \cos^{-1}(|\textbf{B}_{\rm{3D}} \cdot \textbf{B}_{\rm{proj}}|),
\end{equation}
where $\delta \theta_{\rm{proj}} = 0\degree$ again indicates parallel alignment.  Figure \ref{fig:sim_tang_proj} also shows a histogram of $\delta \theta_{\rm{proj}}$ compared to the same statistic derived for two random 3D B-fields.  $\delta \theta_{\rm{proj}}$ has a median value of 33$\degree$.   As with our assumption of tangency, $\delta \theta_{\rm{proj}}$ significantly outperforms a random field ($p < 0.001$). We observe that $\delta \theta_{\rm{proj}}$ generally performs worse than $\delta \theta_{\rm{tang}}$ alone, which is unsurprising given the additional strong assumption introduced.

Several weak-to-moderate correlations exist between $\delta \theta_{\rm{tang}}$, $\delta \theta_{\rm{proj}}$, and properties of the simulated bubble's shell ($n_{\rm{peak}}$, $|B|$, $R_{\rm{SB}}$, $\gamma$), as summarized in Figure \ref{fig:sim_corr}.  A weak negative trend holds between $\delta \theta_{\rm{proj}}$ and shell B-field strength $|B|$ ($\rho_s = -0.25$), indicating slightly improved performance of our combined \AI and \AII with increasing B-field strength.  A weak positive trend holds between $\delta \theta_{\rm{proj}}$ and shell inclination $\gamma$, suggesting that the ability of the magnetized shell to contribute to polarization orientation decreases with increasing $\gamma$.  A weak positive trend holds between $\delta \theta_{\rm{tang}}$ and $n_{\rm{peak}}$, suggesting that the superbubble's B-field is less likely to be tangent to the shell as gas density increases.  We emphasize the overall weak nature of these correlations.

\subsection{Correlations between Bubble Properties and 353 GHz Polarization Metrics}\label{S:sim_corr_planck}

\begin{figure*}
    \centering
    \includegraphics[width=\textwidth]{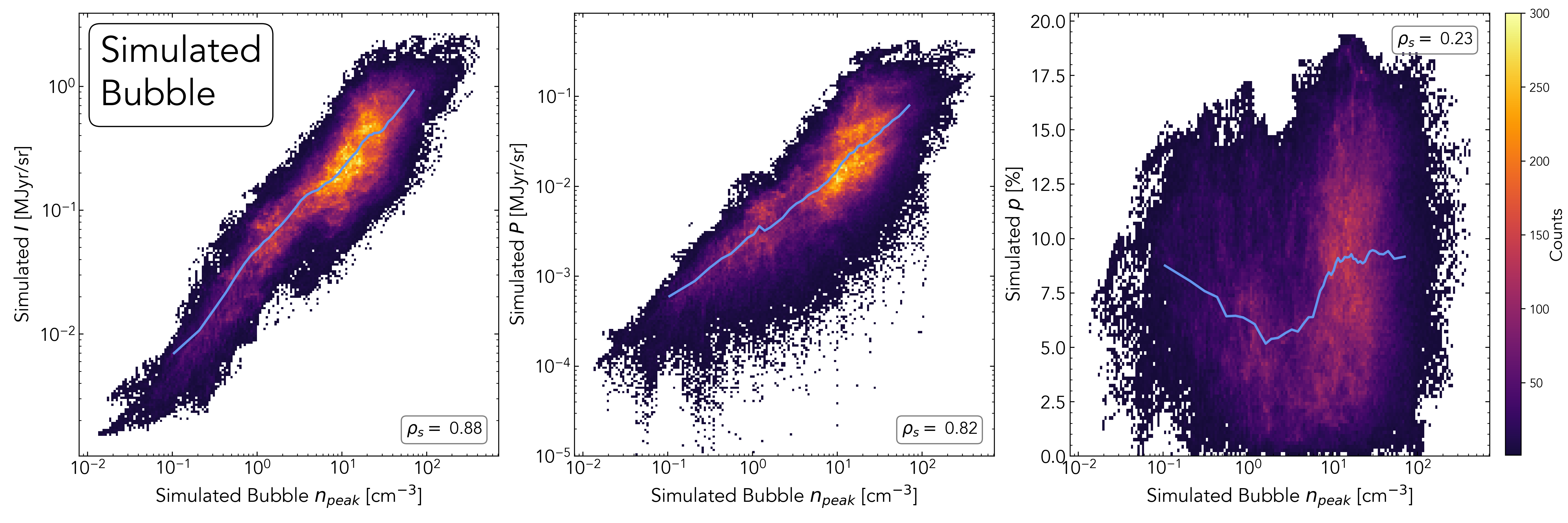}
    \includegraphics[width=\textwidth]{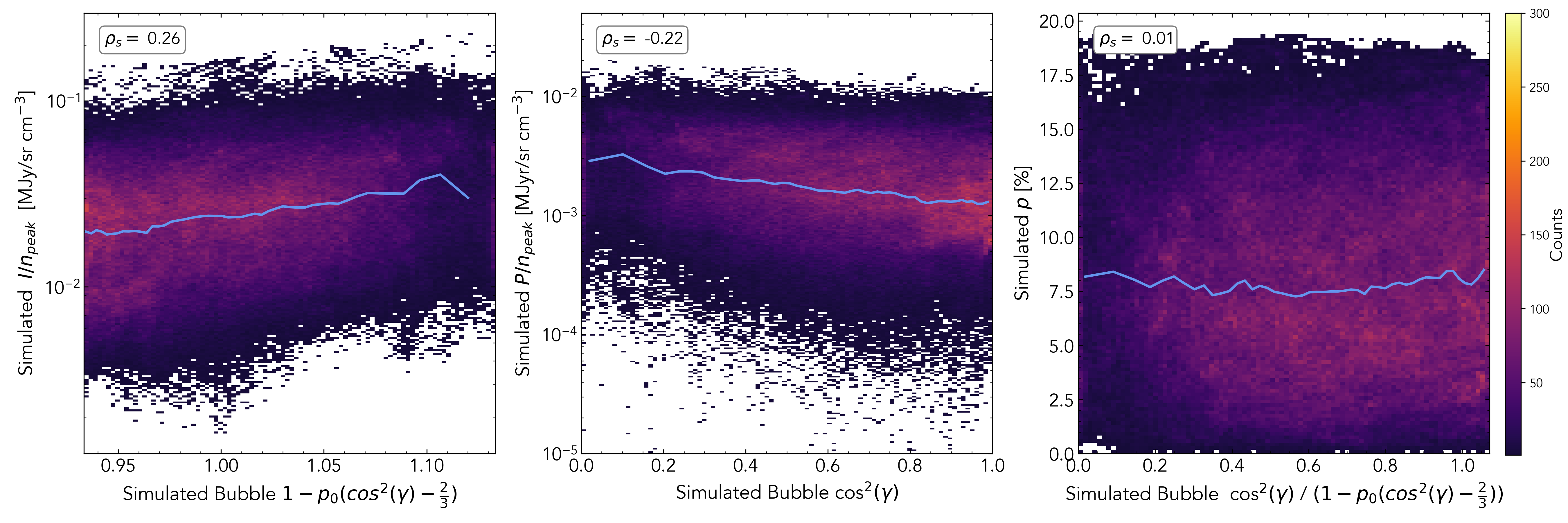}
    \caption{As Figure \ref{fig:planck_corr}, but for correlations between the simulated bubble and simulated 353 GHz polarization metrics.}
    \label{fig:sim_corr_npeak}
\end{figure*}

In \S\ref{S:a1_planck}, we analyzed various correlations between properties of the observed \LB and \planck polarization metrics.  In this section, we reproduce this analysis for the simulated bubble.  As in \S\ref{S:a1_planck}, we restrict our analysis to LOS where the simulated bubble contributes more than 20\% of the total integrated gas density along the LOS ($R_{\rm{SB}} \geq 0.2$, slightly less than the median $R_{\rm{SB}} = 0.23$).  

The top row of Figure \ref{fig:sim_corr_npeak} displays correlations between peak shell density $n_{\rm{peak}}$ and the simulated Stokes $I$, $P$ and $p$ maps; the bottom row displays correlations between shell inclination $\gamma$ and $I/n_{\rm{peak}}$, $P/n_{\rm{peak}}$, and $p$.  Nearly all correlations take on similar trends and strengths to those observed for the real \LB and \planck data.  The exception is $p$ vs. $n_{\rm{peak}}$, which displays a weak positive correlation that is not present in the observed \planck--\LB correlation analysis.  In the subset of simulated LOS where the true 3D B-field's orientation is closest to tangency to the simulated bubble's shell (defined here as being within $5\degree$ of perfect tangency), the lack of correlation between $p$ and the $\gamma$-derived term persists ($\rho_s = 0.04$).

\subsection{Accuracy of Initial B-field Orientation Modeling}

\begin{figure*}
    \centering
      \includegraphics[width=.8\textwidth]{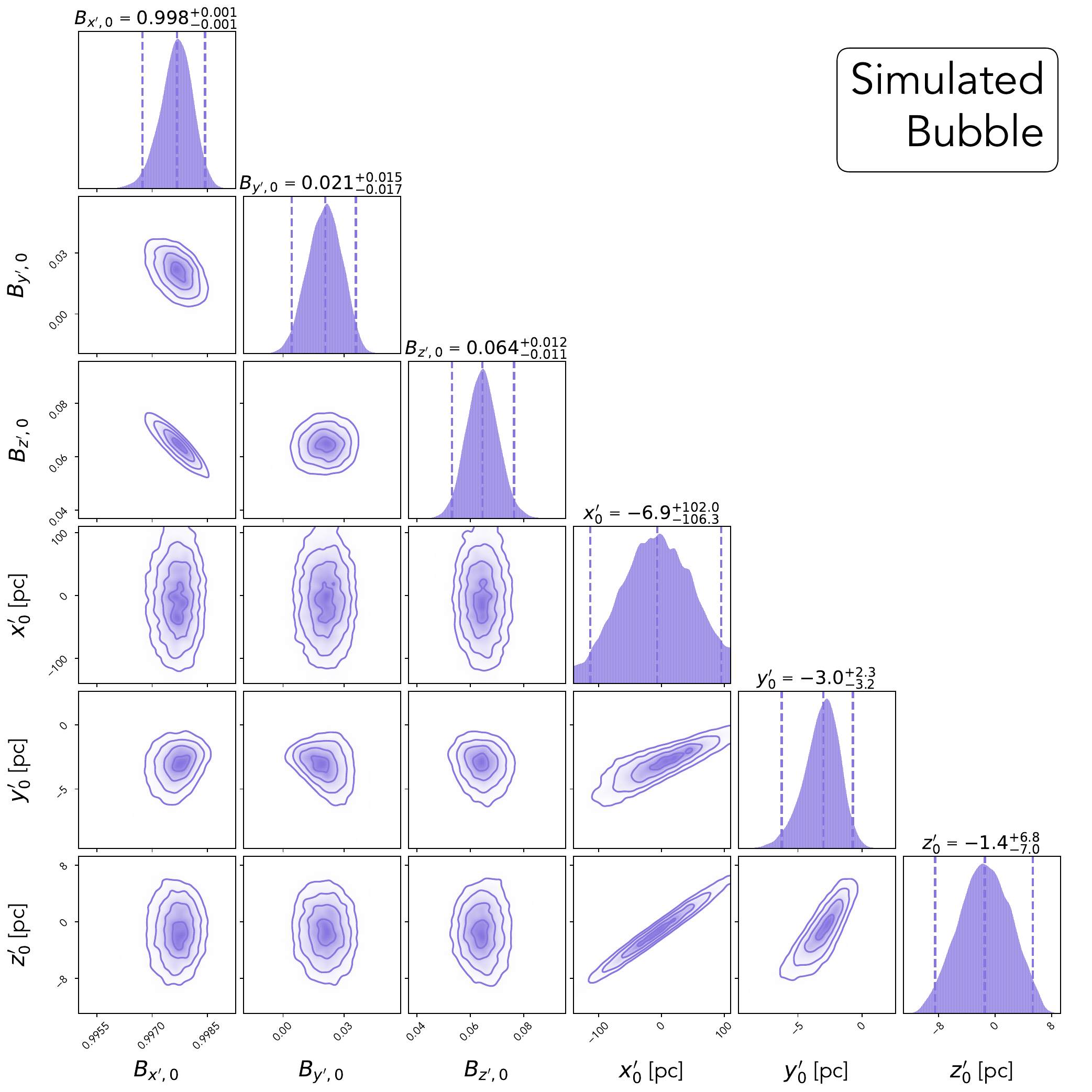}
    \caption{Corner plot showing the posterior distribution of $\theta$ for the simulated bubble.  Reported uncertainties mark the 95\% (2$\sigma$) credible intervals.}
    \label{fig:sim_corner}
\end{figure*}

Finally, we test the accuracy of our ability to reconstruct the initial orientation of the B-field in the ISM surrounding a present-day superbubble, like we do in \S\ref{S:modeling} for the \LB.  The simulated volume containing the M23 superbubble was initialized with a uniform magnetic field oriented in the direction $\textbf{B}_{0}$=($B_{x',0}, B_{y', 0}, B_{z', 0}) =$(1, 0, 0). \footnote{M23 defined a Cartesian coordinate system that we refer to as ($x', y', z'$) where the -$y'$ direction points towards $(\ell, b) = (0\degree, 0\degree)$.  This relates to the Cartesian coordinate system that we use in the bulk of this work (where the +$x$ direction points towards $(\ell, b) = (0\degree, 0\degree)$) as, 
\begin{equation}
\begin{split}
   x'  &= -d \sin (\ell) \cos(b) = -y \\
   y' &= -d \cos(\ell) \cos(b) = -x \\
   z' &= d \sin(b) = z .
\end{split}
\end{equation}
While analyzing the M23 bubble, we make use of the M23 coordinate system.}    

We perform our modeling procedure for the simulated $\textbf{B}_{\rm{proj}}$ largely as described in \S\ref{S:modeling}, with the only change coming in our priors, which are selected to be appropriate for the center and $A_{0.9}$ extent of the simulated bubble:
\begin{equation}
\begin{split}
\textbf{B}_0 &  \begin{cases}
B_{x', 0} \in [0, 1] \\
B_{y', 0} \in [-1, 1] \\
B_{z',0} \in [-1, 1] \\
(B_{x', 0}^2 + B_{y', 0}^2 + B_{z', 0}^2)^{1/2} = 1 \\
\end{cases} \\
c_0 &  \begin{cases}
    x'_0 \sim \mathcal{N}(-3, 62) \in [-140, 110] \\
    y'_0 \sim \mathcal{N}(-1, 62) \in [-180, 130] \\
    z'_0 \sim \mathcal{N}(-13, 69) \in [-230, 90]. 
\end{cases}
\end{split}
\end{equation}
We define $\kappa_B = 1$ for all simulated LOS. 

The corner plot for our inferred posterior is shown in Figure \ref{fig:sim_corner}. Our posterior distribution is centered on values of: 
\begin{equation}
\begin{split}
\hat{\textbf{B}}_0 &  \begin{cases}
B_{x', 0} = 0.998 \pm 0.001\\
B_{y', 0} = 0.021 \upm{0.015}{0.017} \\
B_{z',0}  = 0.064 \upm{0.012}{0.011} \\
\end{cases} \\
\hat{c}_0 &  \begin{cases}
    x'_0 = -6.9\upm{102.0}{106.3} \ \textrm{pc} \\
    y'_0  = -3.0 \upm{2.3}{3.2} \ \textrm{pc} \\
    z'_0  = -1.4 \upm{6.8}{7.0} \ \textrm{pc}, 
\end{cases}
\end{split}
\end{equation}
where reported uncertainties mark the 95\% (2$\sigma$) credible intervals.  The angular difference (cos$^{-1}(\hat{|\textbf{B}}_0 \cdot \textbf{B}_{0}|)$) between our inferred $\hat{\textbf{B}}_0$ and the true $\textbf{B}_0$ is 3.9$\degree$.  We deem this a successful reconstruction of the initial orientation of the simulated volume's B-field.  This suggests that our inference procedure is likely to yield a valid description of the initial orientation of the B-field in the neighborhood on an evolved superbubble like the \LB. 


\section{Fitting Results for Initial Local Galactic Magnetic Field}\label{ap:gmf_model}

\begin{figure*}
    \centering
      \includegraphics[width=.8\textwidth]{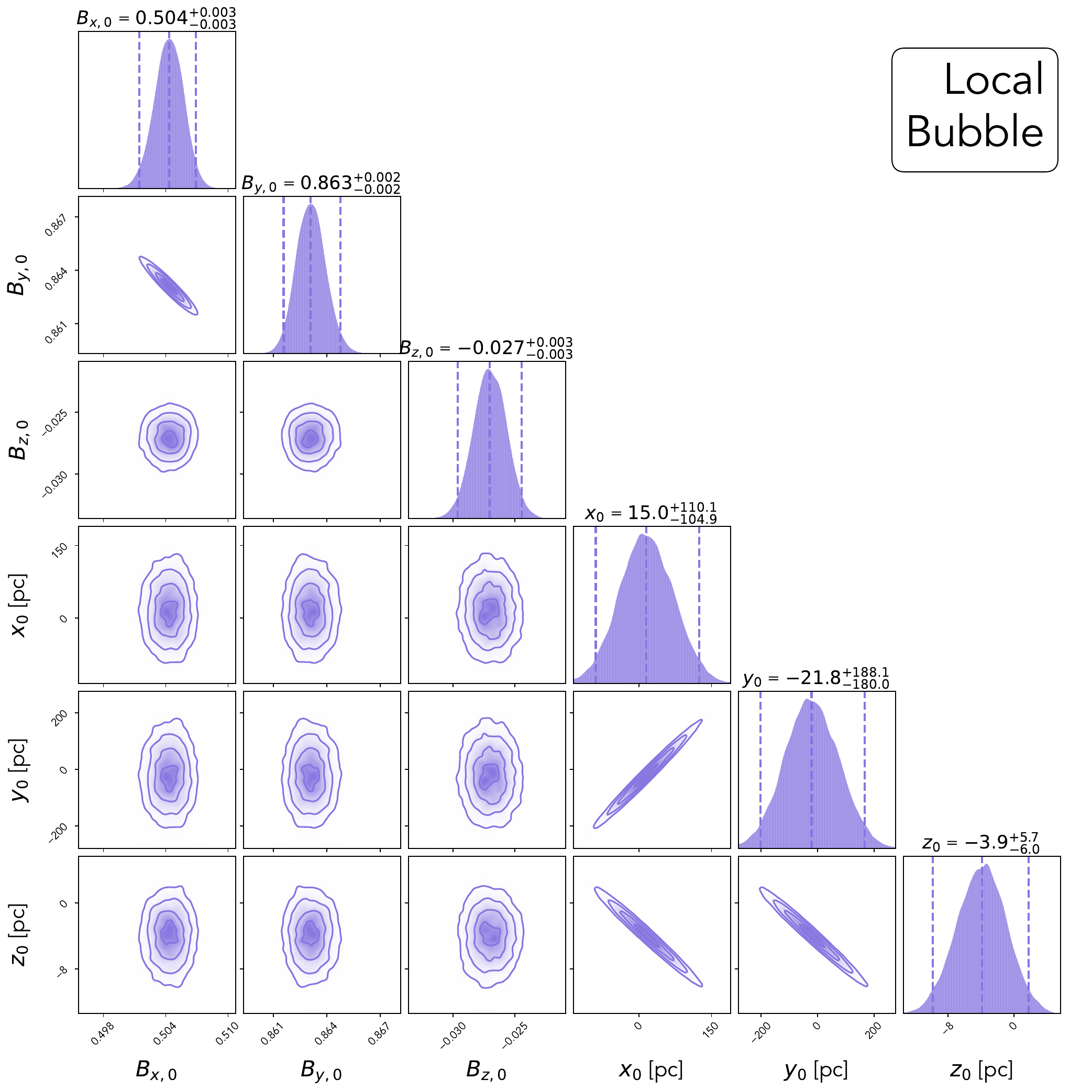}
    \caption{As Figure \ref{fig:sim_corner}, but for $\theta$ sampled for the \LB.}
    \label{fig:gmf_corner}
\end{figure*}

A corner plot showing the posterior distribution of $\theta$ inferred for the observed \LB in \S\ref{S:modeling} is shown in Figure \ref{fig:gmf_corner}.


\bibliographystyle{yahapj}
\bibliography{refs.bib,starpol.bib}

\end{document}